\documentclass[aps,pre,twocolumn,reprint,superscriptaddress,longbibliography]{revtex4-1}
\usepackage{epsfig}
\usepackage{xcolor}
\usepackage{amsmath}
\usepackage{amsfonts}
\usepackage{natbib}
\usepackage{subfigure}
\usepackage{notes2bib}
\usepackage{graphicx}

\newcommand{\kb}{k_{\text{B}}}

\newcommand{\etal}{\emph{et al.}}
\newcommand{\expval}[1]{\langle #1\rangle}

\definecolor{violet}{HTML}{c842fc}

\newcommand{\avg}[1]{\left<#1\right>}
\newcommand{\len}[1]{\left|#1\right|}
\newcommand{\brac}[1]{\left[#1\right]}

\newcommand{\para}[1]{\left(#1\right)}

\newcommand{\kT}{\ensuremath{k_{\rm B}T}}

\newcommand{\Hb}{\ensuremath{\mathcal{H}}}
\newcommand{\hbl}{\ensuremath{\mathbf{h}}}

\newcommand{\Zb}{\ensuremath{\mathcal{Z}}}

\DeclareMathOperator{\real}{Re}

\begin{document}

\title{Frustration and topology in the phase behavior of solute-frustrated $q$-clock models}

\author{Hari Rajesh}
\altaffiliation{Present address: Department of Chemistry, Northwestern University,
2145 Sheridan Road, Evanston, Illinois 60208, USA}
\author{Richard C. Remsing}
\email[]{rick.remsing@rutgers.edu}
\affiliation{Department of Chemistry and Chemical Biology, Rutgers University, Piscataway, NJ 08854, USA}

\begin{abstract}
Frustration can enhance fluctuations and create new and interesting phase behavior. 
When combined with the ability to host topological defects, frustration can increase the complexity of a system's phase diagram through topological and frustrated topological transitions. 
To understand the interplay of frustration and topology on phase behavior, we introduce and study solute-frustrated $q$-clock (SF$q$C) models, in which solute sites are added to the $q$-clock model, a lattice model where each spin can take one of $q$ orientations. 
The $q$-clock model can exhibit spontaneous symmetry breaking and topological transitions, and the interaction between the spins and a solute frustrates the system, in essence by creating a fixed lower bound on the concentration of positive topological defects or vortices. 
We perform Monte Carlo simulations and develop theoretical descriptions of SF$q$C models with $q=4$ and $q=6$ to understand their phase behavior. 
We find that these systems exhibit rich phase diagrams that depend on the choice of parameters. 
In particular, SF$q$C models have a wide range of interesting low-temperature phases, including the presence of strongly frustrated domains and the absence of long-range order. 
We also find topological effects even at low temperatures; for instance, we find that topological defects can drive crystallization. 
We anticipate these results to guide the development of materials that take advantage of frustration and topology to tune phase behavior and fluctuations. 
\end{abstract}

\maketitle

\raggedbottom

\section{Introduction}

The inability of a system to satisfy all of its interactions at the same time --- frustration --- manifests degenerate ground states, enhanced fluctuations, and rich phase diagrams~\cite{toulouse1977,moessner2006}.
In lattice models, frustration is typically a fixed property of the system arising from the lattice geometry as in antiferromagnets on triangular lattices~\cite{wannier1950,chalker1992}, gauge fields as in uniformly frustrated XY models of Josephson junction arrays~\cite{teitel1983}, or quenched disorder as in random field and dilute magnets~\cite{imryma1975,aizenmanwehr1989,cardyostlund1982}.
Here, we consider systems where frustration instead arises from mobile solute particles that interact with one another and spins in the system. 
As a result, these solutes can respond to the surrounding spins and assemble in non-trivial structures, giving rise to rich phase diagrams. 
This type of annealed frustration appears in fluids, for example, where the solute is a physical ion or molecule. 
We are motivated by electrolyte solutions confined to a single monolayer in the interlayer of layered two-dimensional materials~\cite{remsing2015,thenuwara2016,bhullar2021,algarasiller2015}.
In these highly confined environments, the continuous translational symmetry of a fluid is often broken, because the solvent molecules (e.g. water) bind to specific sites on the confining surface and diffusion proceeds through discrete hops between lattice sites~\cite{remsing2015,remsing2017,munoz2021}. 
Moreover, this confining environment can restrict the orientations of the solvent dipoles. 
When solvating ions in this confined space, solvent dipoles point radially outward from a cation, and the resulting frustrated solvation structures can enhance electron transfer rates~\cite{remsing2015,bhullar2021}. 
As a result, these confined electrolytes have unique and tunable electrochemical properties~\cite{chmiola2006,thenuwara2016,bhullar2021}.
A charge-frustrated XY model, a classical XY model with additional ionic sites~\cite{remsing2015,bhullar2021}, captures this frustration-enhanced reactivity. 
Yet, the phase behavior of this class of models has remained unexplored. 
We introduce and study a short range analogue of the charge-frustrated XY model: solute-frustrated $q$-clock (SF$q$C) models, which combine a $q$-state clock model~\cite{jose1977,elitzur1979,tobochnik1982,lapilli2006,ortiz2012,li2020clock} with a lattice gas of charged solutes that repel one another at short distances and interact with spins in a way that orients the spins radially outward from the solute, creating a topological defect or vortex. 
As a result, the solutes create a fixed concentration of solute-centered, mobile positive topological defects in the system; additional defects involving spin-only degrees of freedom can form and annihilate as well. 
Without solutes, a SF$q$C model reduces to the corresponding $q$-clock model, which exhibits a Berezinskii-Kosterlitz-Thouless (BKT) topological phase transition for $q\geq5$~\cite{berezinskii1971,kosterlitz1973,jose1977,elitzur1979}.
If the solutes are uncharged, such that each solute site becomes a vacancy, the model reduces to a dilute $q$-clock or Potts model~\cite{Nienhuis1979,Cardy1979,Berker1979}.
The SF$q$C models are also related to the Coulomb- and charge-frustrated Ising models of microemulsions and ionic liquids~\cite{stillinger1983,wu1992,woo1995,woo1996,dawson1990,deem1994a,deem1994b,grousson2000,grousson2001,limmer2015,bozym2015,ludwig2018}, in which macroscopic phase separation is prevented by the constraints induced by Coulomb interactions. 
The SF$q$C models do not have these electrostatic constraints, because they are short ranged, and we save the study of long ranged charge-frustrated analogues for future work. 
We map out the phase diagrams for the SF4C model on a square lattice and the SF6C model on a hexagonal lattice for small ($Q=1$) and large ($Q=5$) solute charges using Monte Carlo simulations with parallel tempering. 
We uncover rich, charge and $q$-dependent phase diagrams that arise from the combination of frustration and topology.
We rationalize many of these phases through the development of approximate theories and mappings onto simpler models. 
Example phases include solute crystallization, which can be explained by mapping the system onto an antiferromagnetic Ising model with effective interactions prescribed by a dielectric-like theory;
microemulsion-like states in which solutes assemble into one-dimensional walls that divide the spins into cells that exhibit non-trivial topological structure; 
topological crystals that result from ordering of solute-centered vortices and the neutralizing antivortices that form in the spin degrees of freedom; and BKT critical regions that exist at low solute concentrations. 
Many of these phases have analogues in experimentally realizable systems, such that the model presented may be of broad interest beyond the nanoconfined liquids that motivate the work presented here. 

\section{Model and Simulation Methods}

We consider a two-dimensional lattice with size $\ell\times\ell$ of sites $i$ with a total number of sites  $N=\ell^2$. 
A site can be occupied by either a spin or an ion, with this occupation indicated by $t_i$; $t_i=0$ if site $i$ is occupied by a spin and $t_i=1$ for an ion (or solute, we will use the two interchangeably here). 
The state point of the system is then described by the temperature, $N$, and the number density of ions, $\rho = \frac{1}{N}\sum_{i=1}^N t_i$. 
For a $q$-clock model, a spin can adopt one of $q$ orientations, $\theta_i\in\{0,\frac{2\pi}{q},\ldots,2\pi-\frac{2\pi}{q}\}$; we will also often use the vector notation $\mathbf{S}_i=(\cos(\theta_i),\sin(\theta_i))$ to describe the spin orientation.
We write the Hamiltonian of our model as a sum of spin-spin, spin-ion, and ion-ion interactions,
\begin{equation}
\mathcal{H}=\mathcal{H}^{\text{SS}}+\mathcal{H}^{\text{SI}}+\mathcal{H}^{\text{II}}.
\end{equation}
The first term of $\Hb$ is the spin-spin interaction, which is identical to that of the unfrustrated $q$-clock or XY models,
\begin{equation}
\mathcal{H}^{\text{SS}}=-J\sum_{\expval{i,j}}(1-t_i)(1-t_j)\cos(\theta_i-\theta_j),
\end{equation}
where $J$ is the spin-spin coupling constant and the notation $\expval{i,j}$ indicates that the sum is performed over nearest-neighbors. 
The spin-ion interaction is given by 
\begin{equation}
\mathcal{H}^{\text{SI}}=-KQ\sum_{\expval{i,j}}(1-t_i)t_j\cos(\theta_i-\phi_{ij}),
\end{equation}
where $\phi_{ij}$ is the angle between $\mathbf{r}_{ij}$, the vector connecting sites $i$ and $j$, and the $x$-axis, $Q$ is the charge on the ion (here taken to be the same for all ions), and $K$ is the spin-ion coupling constant. 
The spin-ion interaction favors spins pointing radially outward from an ion for $K>0$. 
The last term in $\Hb$ is the ion-ion interaction, $\Hb^{\text{II}}$.
In its most general form, the ion-ion interaction can be written as
\begin{equation}
\mathcal{H}^{\text{II}}=\frac{Q^2}{2}\sum_{i,j}t_it_jV(r_{ij}),
\end{equation}
where $V(r_{ij})$ is the potential at site $i$ from a charge at site $j$. 
For simplicity, we choose a simple short-range, nearest-neighbor repulsion for $V(r)$,
\begin{equation}
\mathcal{H}^{\text{II}}=LQ^2\sum_{\avg{i,j}}t_it_j,
\end{equation}
where $L$ is the ion-ion coupling constant. 
The use of a short-range $V(r)$ enables us to focus on non-trivial phase behavior produced by local frustration, akin to other locally frustrated systems like the ANNNI model. 
However, for true ionic interactions, $V(r)$ is Coulombic and long-range. 
When $V(r)$ is long-range, macroscopic phase separation of the charged species is forbidden due to the constraint of electroneutrality. 
Instead, such charge- or Coulomb-frustrated models exhibit microphase separation~\cite{grousson2000,grousson2001,deem1994a}. 
We reserve the study of long-range variants of our models (charge-frustrated $q$-clock models) for future work, and here focus on short-range \emph{solute}-frustrated $q$-clock models.
We limit our study to two clock models: the SF4C ($q=4$) and SF6C ($q=6$) models, which are sufficient to illustrate the richness of the phase behavior in these systems. 
For a two-dimensional $q$-clock model ($\rho=0$), topological BKT transitions exist for $q>5$ but not for $q<5$, such that our choice of $q$-values straddles this transition in $q$. 
To maintain rotational isotropy with the addition of ions, we study the SF4C model on a square lattice with $z=4$ nearest neighbors per site, and we study the SF6C model on a hexagonal lattice with $z=6$. 
For both models, we investigate $Q=+1$ and $Q=+5$ charges and set $J=K=L=1$. 
Furthermore, we performed simulations of each model at 36 different ionic concentrations, $\rho$, evenly spaced between zero and one, excluding $\rho=1$, because a system with only ions is trivial. 
We performed Metropolis Monte Carlo simulations of each model in the canonical (NVT) ensemble with parallel tempering. 
We kept the number of ions fixed during the simulation. 
At each time step, we considered two possible moves, performed one after the other. 
First, we considered rotating a spin, as in the unfrustrated $q$-clock model. 
In the second, we considered swapping the positions of two particles---this move allows the ions to ``diffuse'' throughout the system, in addition to changing spin orientations. 
We use parallel tempering to adequately sample the many free energy basins in our models.
The total number of temperatures simulated, $N_T$, as well as the minimum and maximum temperatures simulated for each system, $T_{\rm min}$ and $T_{\rm max}$, are listed in Table~\ref{table:temps}.
The temperature range was broken up into smaller ranges of 40 simulations each; the intermediate temperatures $T'$ and, for the $Q=+5$ systems, $T''$ are also listed in Table~\ref{table:temps}.
Temperatures were chosen in a geometric ratio such that within each smaller range, $\frac{T_{i+1}}{T_i}$ is constant. 
This is only optimal in the case of a constant heat capacity~\cite{earldeem2005}, but was sufficient for our purposes.
For a system size of $24\times 24$, we ran the system for $10^{10}$ total time steps, of which the first $9\times 10^9$ steps were taken as equilibration. 
We performed parallel tempering moves every $10^4$ time steps. 
For systems with $Q=+1$, we performed two sets of simulations for each run, one at low temperatures and one at high temperatures, while for systems with $Q=+5$, we performed three sets of simulations. 
The specific temperatures chosen are shown in Table \ref{table:temps}. 
We estimate error bars from the standard deviation across five independent simulations for each system.
To investigate finite size effects, we performed additional simulations of select systems on a $48\times48$ lattice for $10^{11}$ time steps, $9\times 10^{10}$ of which were taken as equilibration.
We did not carry out an in depth finite size analysis for all systems due to the large number of simulations required and the significant increase in simulation time needed to adequately sample configurations for the larger lattices. 
%

\begin{table}
\centering
\begin{tabular}{|c|c|c|c|c|c|c|} 
 \hline
 System & $Q$ & $T_{\text{min}}$ & $T'$ & $T''$ & $T_{\text{max}}$ & $N_T$ \\
 \hline
 SF4C & +1 & 0.04 & 2 & - & 10 & 80 \\ 
 \hline
 SF4C & +5 & 0.04 & 3 & 8 & 20 & 120 \\
 \hline
 SF6C & +1 & 0.04 & 2 & - & 10 & 80 \\
 \hline
 SF6C & +5 & 0.04 & 3 & 8 & 20 & 120 \\
 \hline
\end{tabular}
\caption{
Minimum and maximum temperatures studied, as well as the number of temperatures $N_T$. 
All temperatures are in units of $\frac{J}{\kb}$. 
}
\label{table:temps}
\end{table}

\section{Phase Diagrams}

\begin{figure}[t]
\begin{center}
\includegraphics[width=0.45\textwidth]{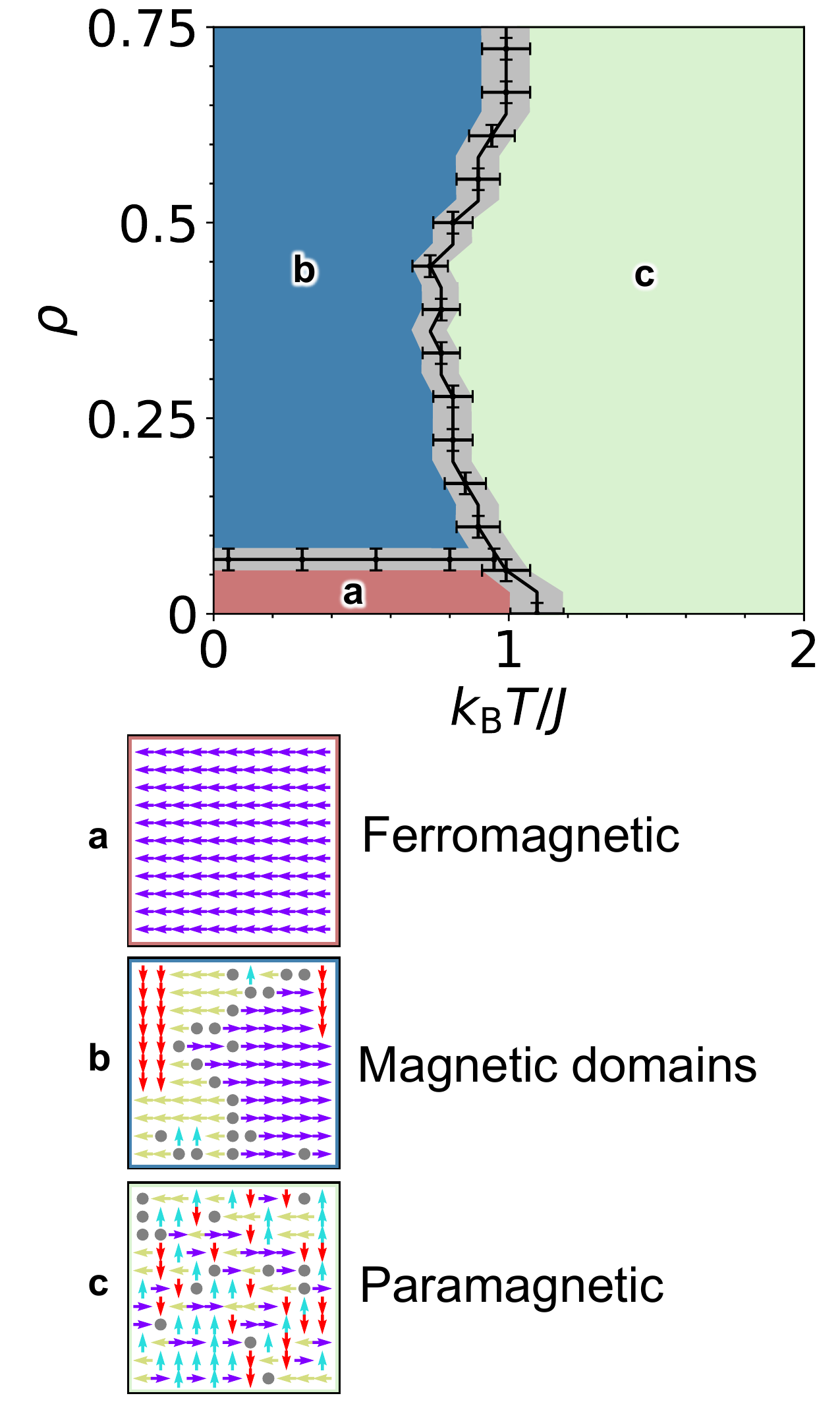}
\end{center}
\caption{
Phase diagram of the SF4C model with $Q=+1$. 
}
\label{fig:sf4c_q1_phases}
\end{figure}

\begin{figure}[h]
\begin{center}
\includegraphics[width=0.5\textwidth]{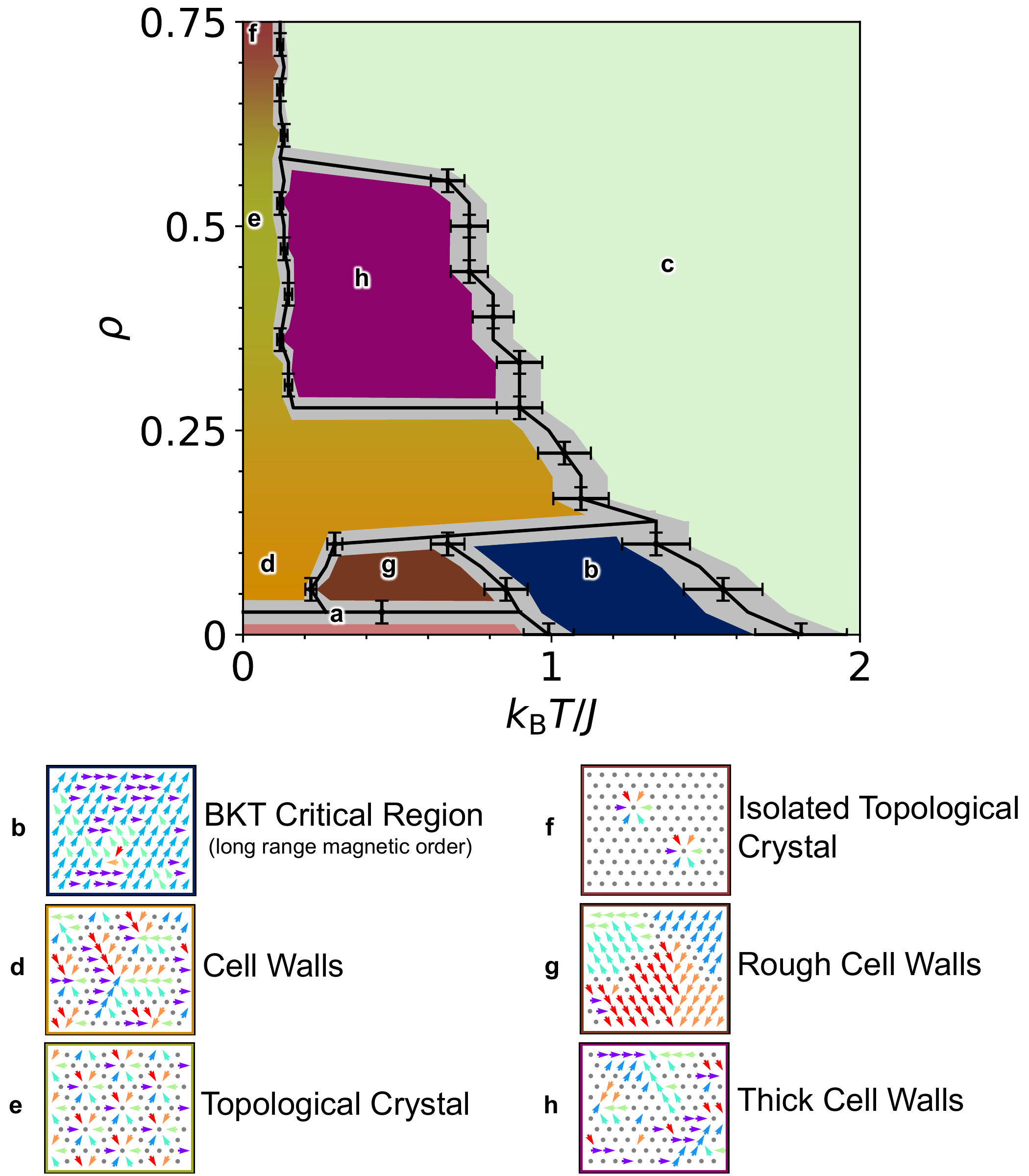}
\end{center}
\caption{
Phase diagram of the SF6C model with $Q=+1$. 
}
\label{fig:sf6c_q1_phases}
\end{figure}

\begin{figure*}[tbh]
\begin{center}
\includegraphics[width=0.9\textwidth]{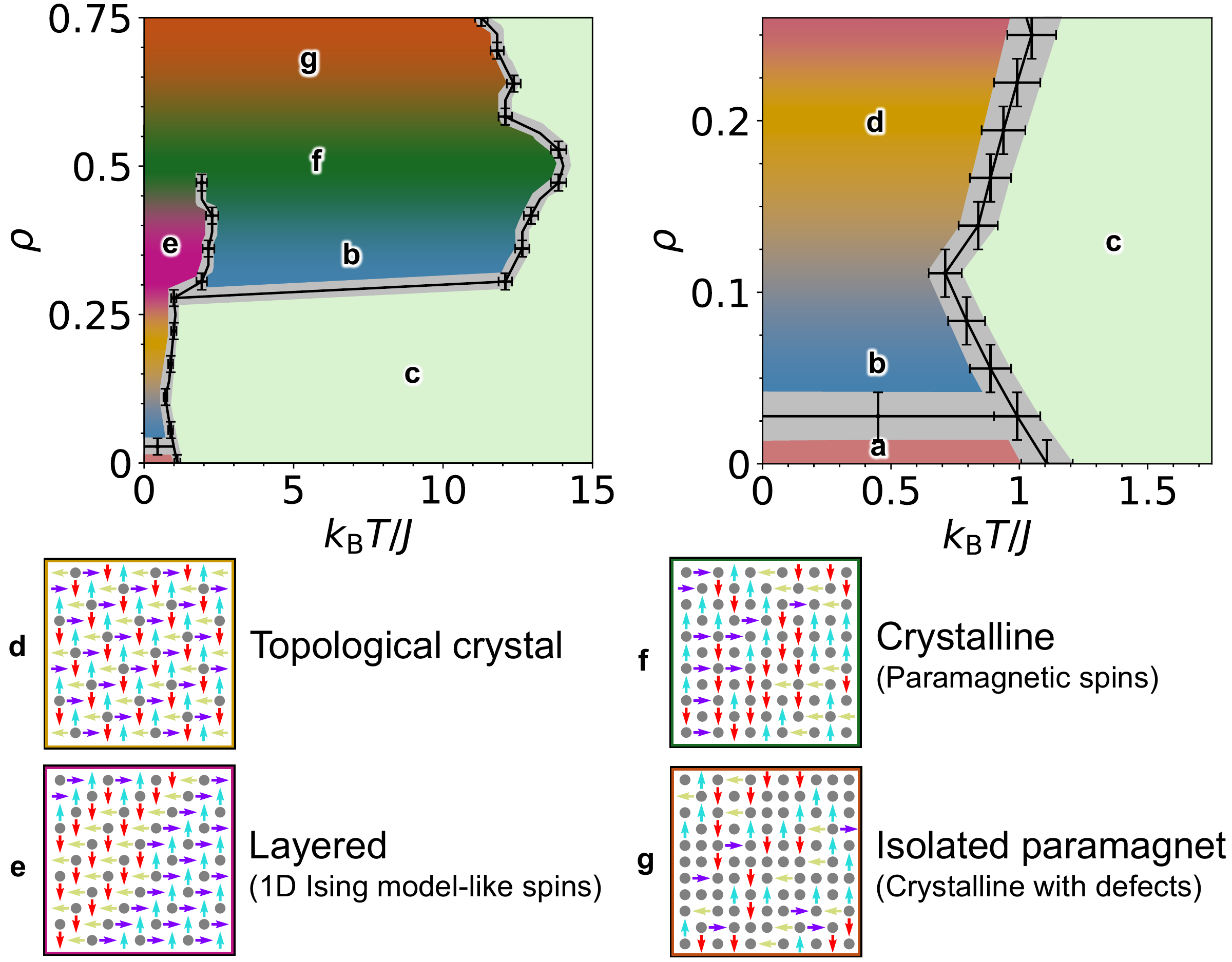}
\end{center}
\caption{
Phase diagram of the SF4C model with $Q=+5$. 
}
\label{fig:sf4c_q5_phases}
\end{figure*}

\begin{figure*}[tbh]
\begin{center}
\includegraphics[width=0.9\textwidth]{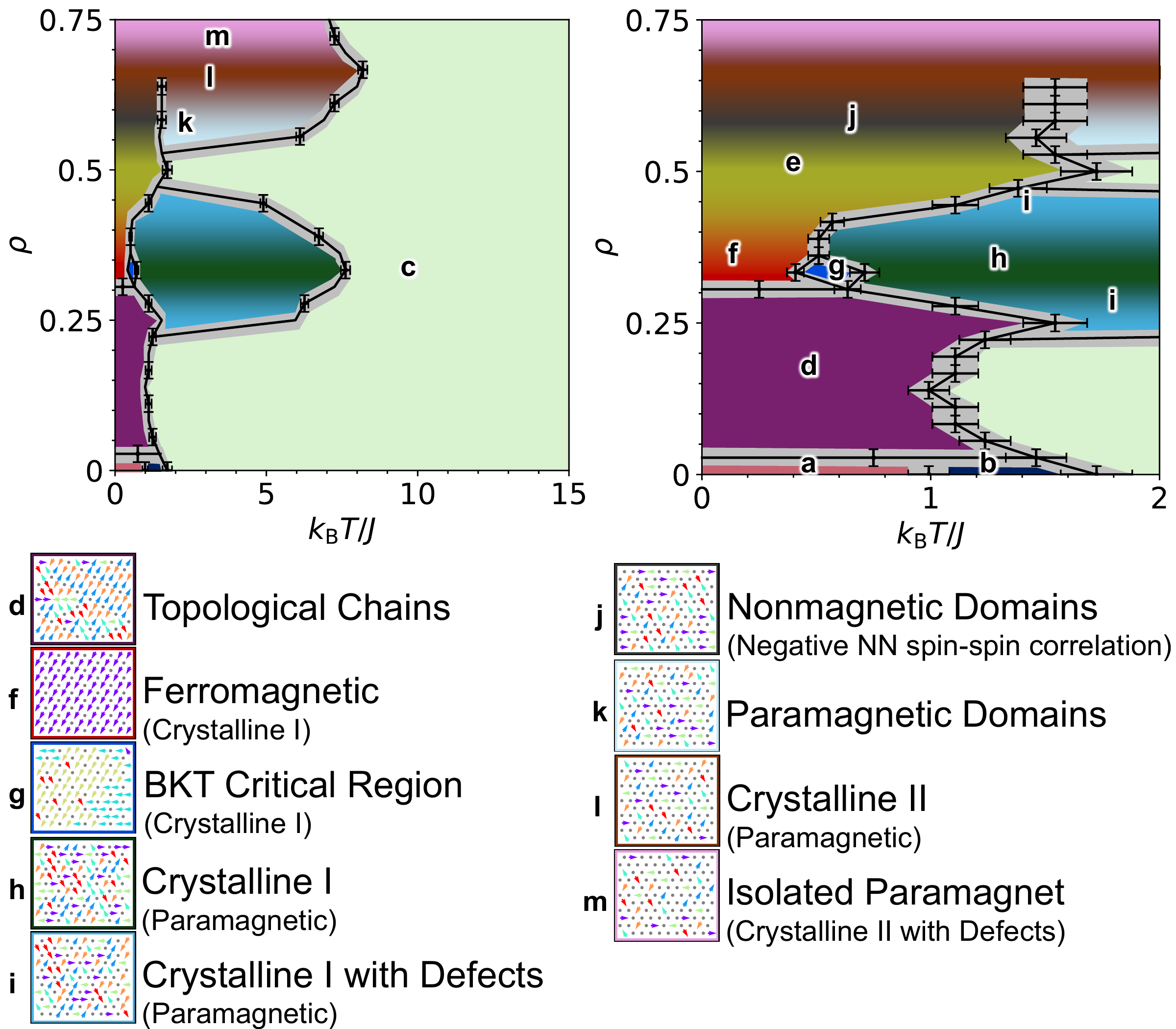}
\end{center}
\caption{
Phase diagram of the SF6C model with $Q=+5$. 
}
\label{fig:sf6c_q5_phases}
\end{figure*}

In Figures \ref{fig:sf4c_q1_phases}-\ref{fig:sf6c_q5_phases}, we show the phase diagrams of the SF4C and SF6C models for $Q=+1$ and $Q=+5$. 
These diagrams were created semiquantitatively, using the total heat capacity, ion-ion heat capacity, magnetic susceptibility, and various order parameters, including the magnetization and others defined later in the paper, to estimate transition temperatures.
The transition temperatures are estimated for the finite size $24\times24$ systems; we reserve the full finite size analysis required to estimate transition temperatures in the thermodynamic limit for future work.
The width of the gray lines and the accompanying error bars indicate uncertainty in the location of the phase transition.
Regions in which phase coexistence and phase separation occur are indicated by smooth color gradients. 
%

%
We now analyze the phase behavior of SF$q$C models. 
We break this analysis into two sections. 
In Section~\ref{sec:nontopphases}, we focus on phases that are non-topological. 
Then, in Section~\ref{sec:topphases}, we focus on phases that are topological in nature. 

\section{Non-Topological Phases}
\label{sec:nontopphases}

\subsection{Ionic Crystallization}

\begin{figure}
\begin{center}
\includegraphics[width=0.45\textwidth]{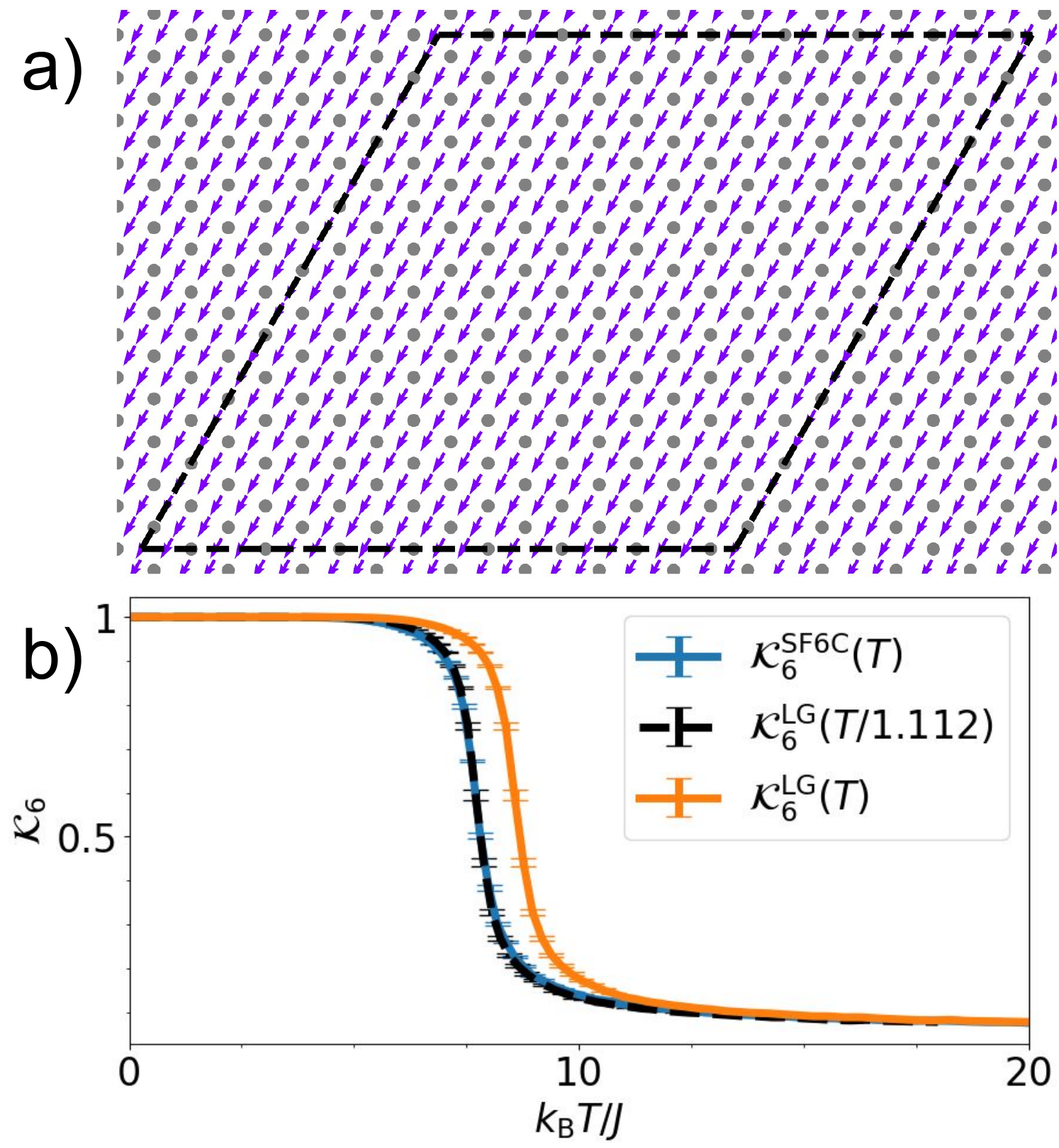}
\end{center}
\caption{
(a) Low-temperature configuration of the SF6C model with $Q=+5$ ions at a concentration of $\rho=\frac{1}{3}.$ 
Note that the black dashed lines indicate the periodic boundaries. 
(b) Comparison of $\mathcal{K}_6$ between the SF6C model, a lattice gas, and the lattice gas at a shifted temperature scale, with $\rho=\frac{1}{3}$ and $Q=+5$ for both 
}
\label{fig:crystop}
\end{figure}

First, we note that in the highly charged $Q=+5$ case, all systems with ions undergo a transition involving ions. 
In particular, all such systems exhibit a peak in the ion-ion heat capacity, below which the ion-ion energy is roughly minimized. 
This peak is typically quite diffuse, except at specific values of $\rho$ where the peak sharpens.
In particular, this occurs at $\rho=\frac{1}{2}$ for the SF4C model and at $\rho=\frac{1}{3},\frac{2}{3}$ for the SF6C model, and we refer to these as crystalline phases.
In the SF6C model, the crystal at $\rho=\frac{1}{3}$ isolates ions from one another, such that each ion is completely surrounded by spins and spin-spin contacts persist as well. 
At $\rho=\frac{2}{3}$, the resulting crystal is the reverse; each spin is completely surrounded by ions and ion-ion contacts persist. 
In all three of these cases, there is a single unique ground state for the ions whereby moving any ion will increase the energy. 
The ion concentrations at which these crystalline phases can occur is dictated by the geometry of lattice, with $\rho$ of a crystalline phase equivalent to that which enables each spin to be isolated from the other spins or each ion to be isolated from the other ions. 
For the square lattice, isolation of spins and ions occur at a single concentration, $\rho=1/2$. 
For the hexagonal lattice, isolation of ionic sites occurs at $\rho=1/3$ and isolation of single spins occurs at $\rho=2/3$, leading to the Crystalline~I and Crystalline~II phases, respectively.
The transition to crystalline phases can be understood by mapping the SF$q$C models to lattice gas and antiferromagnetic Ising models. 
As shown in Appendix~\ref{sec:mapping}1, we can derive an approximate effective Hamiltonian by integrating out the spins,
\begin{equation}
\Hb_{\rm eff} = U_{\rm eff}(T) N_{\rm II},
\end{equation}
where $N_{\rm II}=\sum_{\avg{i,j}}t_i t_j$ is the number of ion-ion contacts in a configuration.
The effective interaction is
\begin{equation}
U_{\rm eff}(T) = \frac{U_0}{\varepsilon(T)},
\end{equation}
where the bare ion-ion interaction is given by the Hamiltonian without spins, $U_0=LQ^2$, and the effective dielectric constant that screens the ion-ion interactions is
\begin{equation}
\varepsilon(T) = \para{1-\frac{K^2 a}{4L \kT}}^{-1} \sim 1+\frac{K^2 a}{4L\kT},
\end{equation}
and $a$ is what we refer to as the number of unbalanced spins per ion-ion contact. 
An unbalanced spin is a spin that has at least one spin and one ion nearest neighbor and the neighbor spins do not cancel each other upon summing over nearest neighbors. 
The form of the dielectric constant is consistent with a Curie-Debye dielectric, $\varepsilon \sim 1/T$, appropriate for noninteracting or independent dipoles.
We can transform this model into an Ising model in the grand ensemble. 
Specifically, if we let $N=\sum_i t_i$ and $\mu$ be the chemical potential, then 
\begin{equation}
\mathcal{H}_{\rm eff}-\mu N=E_0-H\sum_{i}S_i-J_{\rm eff}\sum_{\langle i,j\rangle}S_iS_j,
\end{equation}
which is an Ising model Hamiltonian where $E_0=-\frac{\ell^2\mu}{2}+\frac{U_{\rm eff}\ell^2z}{8}$ is an ignorable zero-point energy, $H=\frac{\mu}{2}-\frac{U_{\rm eff}z}{4}$ is an external magnetic field, $J_{\rm eff}=-\frac{U_{\rm eff}}{4}$ is the spin-spin coupling constant, and $z$ is the coordination number~\cite{leeyang1952}. 
Because $U_{\rm eff}>0$, $J_{\rm eff}<0$, and this model represents an antiferromagnetic (AFM) Ising model. 
The critical behavior of these models is known, such that we can predict how temperature dependent properties will be shifted by the presence of spins that screen the ion-ion interactions,
\begin{equation}
\kT_c^{\rm eff} \approx \kT_c^{(0)} - \frac{aK^2}{4L}
\end{equation}
and $J/J_{\rm eff}=\varepsilon(T)$, as derived in Appendix~\ref{sec:mapping}1 where $T_c^{(0}=cLQ^2$ is the critical temperature of the bare (ion only) lattice gas model, and $c$ is a constant that depends on the lattice geometry and composition. 
The mapping between the SF$q$C model and the AFM Ising model allows us to predict some of the properties of the crystal. 
On a square lattice, we find alternating up and down spins in the AFM Ising model, corresponding with to the crystal observed in the SF4C model that consists of alternating ionic and spin sites. 
On a hexagonal lattice, the extreme degeneracy in a typical AFM Ising model~\cite{wannier1950} is broken by the external field, leading to two solutions. 
In one solution, up spins form a hexagonal lattice and down spins form a honeycomb lattice; in the solution, these roles are reversed. 
The same conceptual framework applies to the ionic crystals in the SF6C model; $\rho=1/3$ shown in Fig.~\ref{fig:crystop}a
This mapping to the Ising model will be useful later when we discuss certain topological phases in the model. 
%

%
We also develop order parameters to quantify ionic crystallization in the SF$q$C models, one for each symmetry type. 
For four-fold symmetry, we define 
\begin{equation}
\mathcal{K}_4=\frac{1}{\rho N}\sum_{i=1}^N t_i e^{\pi i(\mathbf{\hat{x}}+\mathbf{\hat{y}})\cdot\mathbf{r}_i},
\end{equation}
where $t_i$ is 0 for an empty site and 1 for an ionic site, and $\mathbf{r}_i$ is the position of site $i$. 
This order parameter is translationally invariant up to a phase such that $\mathcal{K}'_4=\mathcal{K}_4e^{i\pi\Delta\mathbf{r}\cdot(\mathbf{x}+\mathbf{y})}$; that is, changing the reference position does not change the magnitude of the order parameter. 
For the ordered phase, we get $|\mathcal{K}_4|>0$, while for the disordered phase, we get $|\mathcal{K}_4|=0$. 
Similarly, for six-fold symmetry, we can define 
\begin{equation}
\mathcal{K}_6=\frac{1}{\rho N}\sum_{i=1}^N t_i e^{\frac{2\pi i}{3} (2\mathbf{\hat{x}}+\mathbf{\hat{y}})\cdot\mathbf{r}_i}.
\end{equation}
For $\rho=\frac{2}{3}$, we can use a very similar parameter 
\begin{equation}
\mathcal{K}'_6=\frac{1}{\rho N}\sum_{i=1}^N (1-t_i) e^{\frac{2\pi i}{3} (2\mathbf{\hat{x}}+\mathbf{\hat{y}})\cdot\mathbf{r}_i}.
\end{equation}
In essence, this swaps the roles of spins and ions and measures the crystallization of spins instead. 
These order parameters are also translationally invariant up to a phase, with $|\mathcal{K}_6|>0$ for the ordered phase and $|\mathcal{K}_6|=0$ for the disordered phase. 
Finally, we can compare the results from the pure lattice gas to the SF$q$C models. 
We find that the results are qualitatively and quantitatively similar; we see a shift in the temperature scale at which the transition takes place, an example of which is shown in Fig.~\ref{fig:crystop}b. 
Using this, we can define an effective charge for the ions in the SF$q$C model. 
We note that $T_c\propto Q^2$ in the lattice gas, so the ratio of the transition temperatures can give us the ratio of the effective charge to the true charge. 
From our dielectric theory in Appendix~\ref{sec:mapping}1, we find that $Q_{\rm eff}(T_c)\sim Q - \frac{aK^2}{8cLQ}$.
We also more precisely evaluate the effective charge through an ``isolated system approximation,'' Appendix~\ref{sec:mapping}2, where we determine the effective charge from the square root of the free energy of an isolated portion of the lattice at the critical temperature. 
This serves as a more exact determination of $Q_{\rm eff}$, and both theories are in excellent agreement near the critical point and at high temperatures, Appendix~\ref{sec:mapping}2. 
We additionally determine the ratio of critical temperatures, and therefore $Q_{\rm eff}$, by simply shifting one of the two order parameter curves, which is shown in Fig.~\ref{fig:crystop}b. 
We find that in the SF4C model with $\rho=\frac{1}{2}$, $Q_{\text{eff}}\approx 4.927$, and in the SF6C model with $\rho=\frac{1}{3}$, $Q_{\text{eff}}\approx 4.742$ while for $\rho=\frac{2}{3}$, $Q_{\text{eff}}\approx 4.888$. 
These values of the effective charge are in good agreement with those prediction by the theories, Appendix~\ref{sec:mapping}, and the reduction of the charge is consistent with screening of the ions by neighboring spins, similar to a dielectric fluid~\cite{fumagalli2018,duignan2016,duignan2021,ludwig2018}.

\subsection{Paramagnetic and Ferromagnetic Phases}

We now briefly discuss the simplest magnetic phases in the model. 
Naturally, at high temperatures, in all of the systems, we have a paramagnetic phase where spins are oriented nearly randomly. 
As we shall explore later, however, there is a subtlety here which must be accounted for. 
In these systems, we also find a variety of low-temperature phases which also seem to be paramagnetic in nature from the magnetization alone, and 
we discuss the differences in the phases later.  
There is also a special case of the paramagnetic phase: for certain systems with high amounts of ions, we find that spins are effectively isolated from one another and thus, the system effectively acts as a set of non-interacting paramagnets. 
We will explore this phase in the next section. 
Only a small subset of systems studied demonstrate a true ferromagnetic phase. 
For the SF4C model with $Q=+1$, the addition of a small amount of ions ostensibly allows for a ferromagnetic phase, as evidenced by the magnetization, but this is likely a finite size effect.
This does not appear to be the case in any of the other systems---that is, a small ionic perturbation destroys global ordering. 
The destruction of global order by the presence of ionic sites is consistent with the Imry-Ma argument that long-range order is forbidden in two dimensions for arbitrarily weak disorder~\cite{imryma1975,aizenmanwehr1989}, with the disorder here stemming from the quenched ion positions and the resulting spin-ion coupling.
The ionic sites must be disordered to suppress global order in the system --- ordered arrangements of ions allow long-range order. 
For example, the SF6C model with $Q=+5$ exhibits a ferromagnetic phase at $\rho=1/3$, when the ionic sites crystallize into a hexagonal lattice, Fig.~\ref{fig:crystop}a. 
In this case, the spin-ion interactions cancel at every site due to symmetry, and the spins act like an unfrustrated 6-clock model on a honeycomb lattice and can undergo an order-disorder transition involving a ferromagnetic phase.

\subsection{Isolated Paramagnetic Phase}

\begin{figure}
\begin{center}
\includegraphics[width=0.45\textwidth]{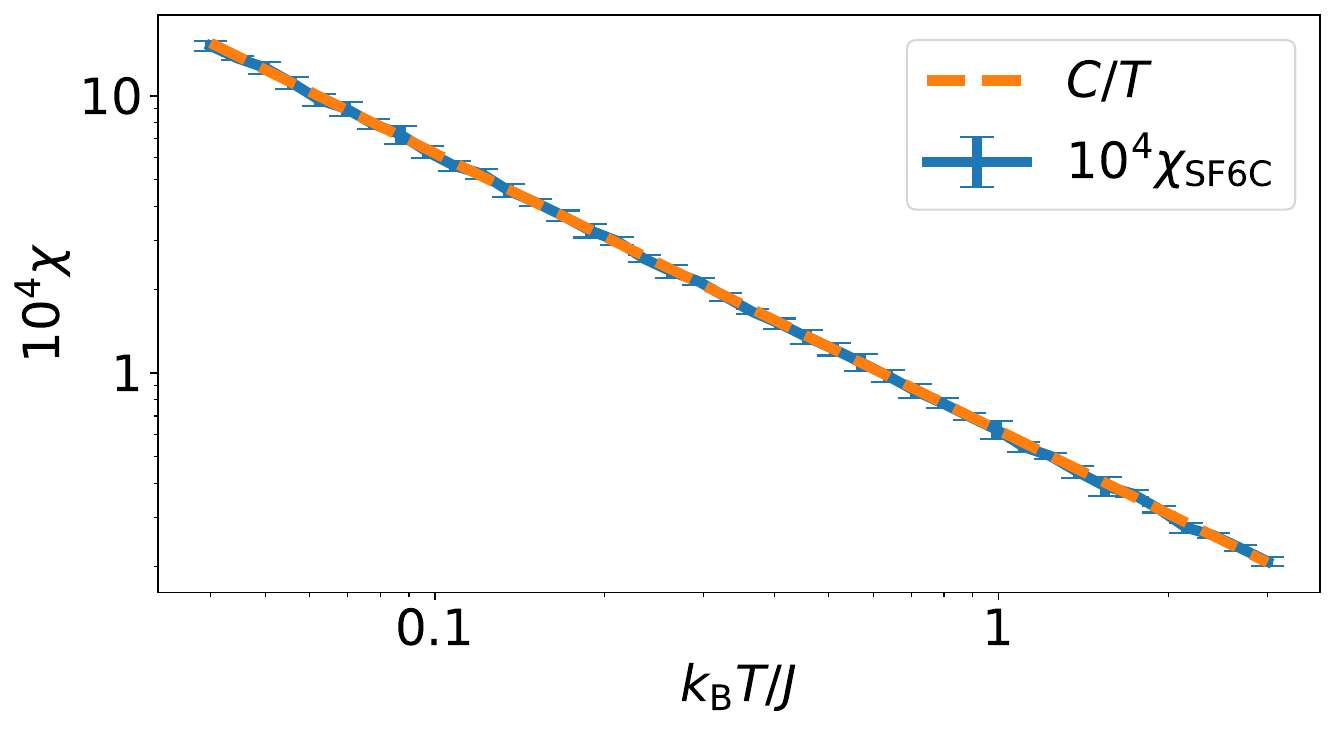}
\end{center}
\caption{
Magnetic susceptibility $\chi$ for the SF6C model at $\rho=\frac{2}{3}$ compared with a $\frac{1}{T}$ curve. 
}
\label{fig:suscep}
\end{figure}

For the highly charged systems, there is no magnetically ordered phase at high ion concentrations, and each system adopts an isolated paramagnetic phase, in which spins are isolated from one another. 
The isolation of spins, or equivalently complete solvation of spins by ions, is entropically favored due to mixing and also energetically favored. 
To illustrate the energetic driving force for spins to be isolated from one another at high $\rho$, consider the limit of two spins on an otherwise completely ionic lattice. 
When each spin is isolated, both the spin-spin and spin-ion interactions are zero. 
If we make the two spins neighbors of each other, the number of spin-ion bonds is reduced from $2z$ to $2z-2$, where $z$ is the coordination number, and one spin-spin bond is created. 
Because the total number of bonds is conserved, this also creates one ion-ion bond. 
The total energetic cost of spin pairing is dominated by the $KQ^2$ of the ion-ion interaction at large $Q$, such that the net change in energy upon pairing two spins is unfavorable. 
This argument readily generalizes to larger spin aggregates.
Because spin-pairing is energetically and entropically unfavorable at high $\rho$, the system will tend to isolate its spins if possible. 
This is possible for $\rho\geq \frac{1}{2}$ on a square lattice and for $\rho\geq\frac{2}{3}$ on a hexagonal lattice. 
We demonstrate the existence of these isolated paramagnetic phases by analyzing the magnetic susceptibility,
\begin{equation}
\chi=\frac{\expval{M^2}-\expval{M}^2}{\kb T},
\end{equation}
which quantifies fluctuations in the magnetization $M$, where $\avg{\cdots}$ indicates an ensemble average. 
The magnetization is given by 
\begin{equation}
\mathbf{M}=\frac{1}{\ell^2}\sum_{i=1}^N(1-t_i)(\cos(\theta_i),\sin(\theta_i)),
\end{equation}
and $M\equiv \expval{M}=\expval{\|\mathbf{M}\|}$. 
In the absence of an external field $\chi\propto\frac{1}{T}$ for a system of non-interacting paramagnets in accordance with Curie's law~\cite{ashcroftmermin}, and as may be expected from our dielectric theory, because $\varepsilon(T)\propto \chi$ in this model. 
We find that for $Q=5$ and for the appropriate concentrations, the susceptibility matches this prediction, as shown in Fig.~\ref{fig:suscep}. 
Finally, we note that in the opposite limit, where $Q\ll 1$, then the spin-spin term dominates the energy. 
In this regime, the formation of spin aggregates or bubbles becomes energetically favorable, creating domains of ordered magnetic phase. 
The model approaches the dilute $q$-clock model as $Q\rightarrow0$, and we expect the phase behavior to be qualitatively similar to the corresponding dilute model in this low charge limit~\cite{Nienhuis1979,Cardy1979,Berker1979}.

\subsection{Layered Phase}

\begin{figure}
\begin{center}
\includegraphics[width=0.45\textwidth]{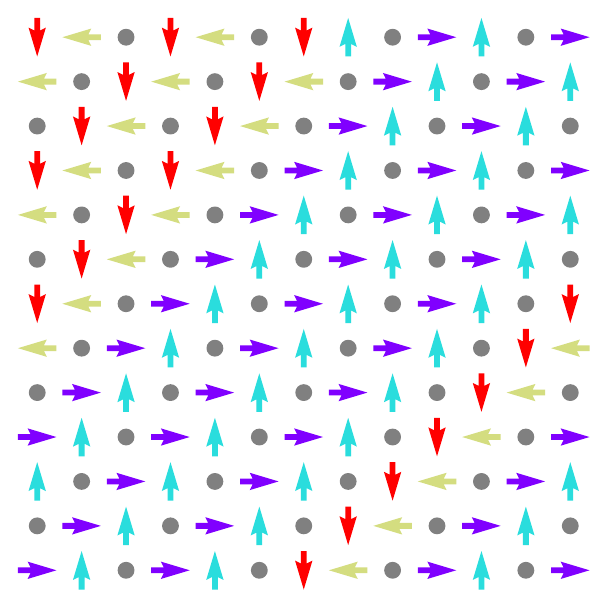}
\end{center}
\caption{
Low-temperature configuration of the SF4C model with $Q=+5$ ions at $\rho=\frac{1}{3}$. 
}
\label{fig:layeredphase}
\end{figure}

The SF4C model with $Q=5$ exhibits a `layered phase' at $\rho=1/3$, Fig.~\ref{fig:layeredphase}.
In the layered phase, ions organize into diagonal sheets, which reduces the (nearest neighbor) ion-ion interactions to zero. 
Each pair of ions within the same plane are separated by a pair of spins.
The orientation of these spins is determined by their interactions with the ions --- the spin-ion coupling at $Q=5$ dominates the spin-spin coupling, and this phase disappears when the two become comparable (e.g. $Q=1$).
However, there is a degeneracy with respect to the orientation of the pair of spins due to frustration, and (moving from the left ion to the right ion) the pair of spins can either be down-left or right-up. 
These arrangements are equivalent and diagonal layers of each pair can nucleate spontaneously. 
Each layer of spin pairs has a net magnetization, but because each layer type is equally probable, the layered phase is on average paramagnetic.
Moreover, we can map each diagonal layer to a one-dimensional ferromagnetic Ising model with the effective exchange coupling equal to half of that in the original model, $J' = \frac{J}{2}$, which suggests that long-range, global order and spontaneous symmetry breaking is not possible within each layer in the thermodynamic limit. 
%

\subsection{Magnetic Domains}

\begin{figure}
\begin{center}
\includegraphics[width=0.45\textwidth]{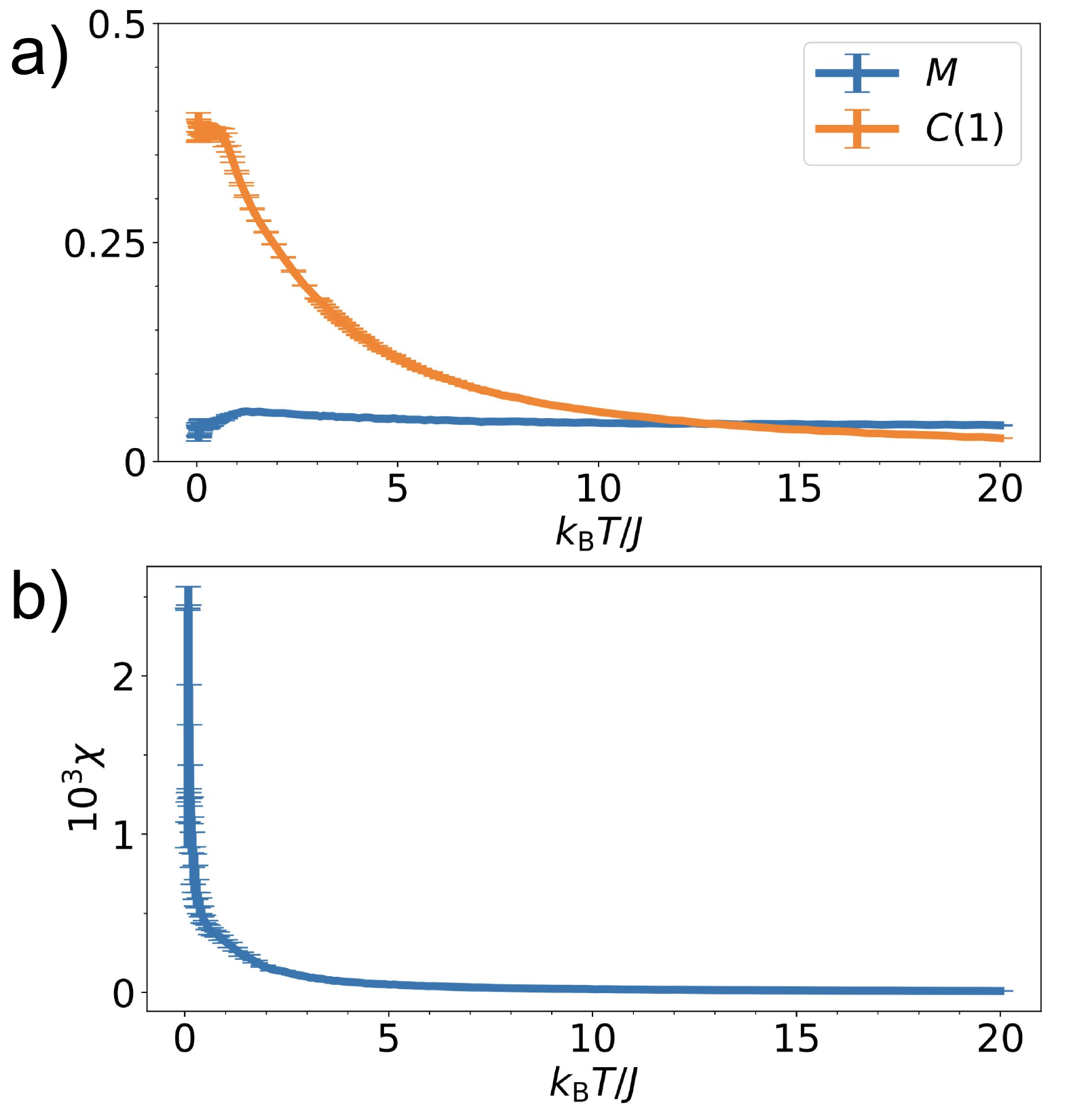}
\end{center}
\caption{
(a) Magnetization and nearest-neighbor spin-spin correlation and (b) magnetic susceptibility in the SF6C model with $\rho=\frac{1}{9}$ and $Q=+5$. 
}
\label{fig:longvsshortorder}
\end{figure}

Any finite concentration of ions frustrates the system, and this frustration prevents global ordering and manifests in the formation of dynamic, fluctuating magnetic domains even at low temperatures. 
We quantify the formation of these domains through comparison of the magnetization and the nearest neighbor spin-spin (NNSS) correlation, $C(1)=\langle \cos(\theta_i-\theta_j)\rangle$, where the average is over nearest neighbor spin pairs $i$ and $j$, Fig.~\ref{fig:longvsshortorder}a. 
As temperature is decreased, $C(1)$ increases, indicating the presence of local spin ordering at low temperatures. 
Despite this local ordering, the magnetization, $M$, remains small at all temperatures and even decreases with $T$ for sufficiently low temperatures. 
This absence of net magnetization despite the prevalence of local order is a hallmark of frustration and domain formation. 
%

\begin{figure}
\begin{center}
\includegraphics[width=0.45\textwidth]{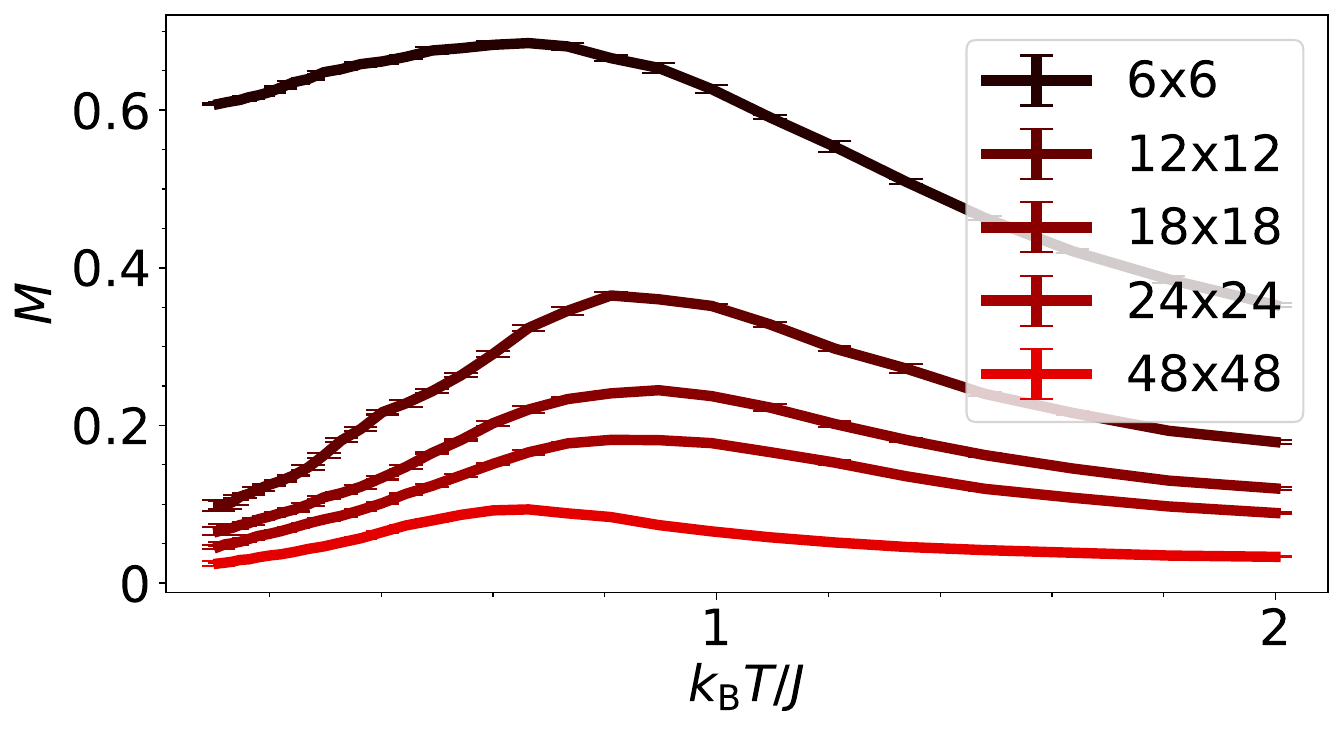}
\end{center}
\caption{
Magnetization of the SF6C $\rho=\frac{1}{4}$ and $Q=+1$ as a function of system size. 
}
\label{fig:sf6cmagsize}
\end{figure}

%
We provide further evidence of magnetic frustration by quantifying the low temperature behavior of the magnetic susceptibility, $\chi\propto \frac{(\Delta M)^2}{T}$. 
For unfrustrated systems, $\Delta M\to 0$ at low temperatures, and $\chi$ remains bounded. 
However, in the presence of frustration, $\Delta M$ never vanishes, and $\chi\to \infty$ as temperature approaches zero. 
Many of the phases we uncover exhibit this behavior, including the isolated paramagnetic phases discussed above, as well as the phases with magnetic domains, as shown in Fig.~\ref{fig:longvsshortorder}b. 
Frustration should also be evident in the system size dependence of the magnetization. 
For small system sizes, local ordering leads to significant magnetization, but as the size increases, formation of multiple domains leads to a reduction in the overall magnetization. 
Indeed, we find that the magnetization strongly decreases as the system size increases, Fig.~\ref{fig:sf6cmagsize}. 
In particular, we find that $M_{\mathrm{peak}}\approx -0.011+3.59\ell^{-0.91}$, such that in the thermodynamic limit, $\ell\to\infty$, we have no net magnetization. 
This is consistent with the idea that the system is ordered over short length scales but disordered over large length scales. 
The size of the magnetic domains is closely connected to the spacing between ions, consistent with theories of microemulsions, as discussed further in Section~\ref{subsec:cells}. 
%

%
We also note that these magnetic fluctuations remain even if the ions are stationary. 
These fluctuations could be due to each ionic configuration having a different yet non-degenerate ground state magnetization. 
To disprove this, we may consider a variant of the SF$q$C model in which the solute sites are placed randomly and held fixed, such that the only allowed Monte Carlo move is a spin flip. 
In this frozen ion model, the magnetic susceptibility still diverges at low temperatures, indicating that the spins are truly frustrated and have no single ground state. 
This lack of a single ground state is consistent with the Imry-Ma argument that long-range order is forbidden in two dimensions for arbitrarily weak disorder~\cite{imryma1975,aizenmanwehr1989}, with the disorder here stemming from the quenched ion positions and the resulting spin-ion coupling.
Consider the limiting case of these magnetic domains when the ionic charge is very large and the concentration approaches the value where we expect the isolated paramagnetic phase. 
In this regime, we have regions with only a few spins surrounded by ions. 
Consequently, the behavior of the spins is dictated by the ions. 
In particular, consider an isolated domain with two spins surrounded by ions. 
Naturally, owing to the spin-ion interaction dominating the spin-spin interaction, we expect the spins to point towards one another. 
As a result, the nearest-neighbor spin-spin correlation in this regime becomes strongly negative. 
We use mean field theory to predict this nearest-neighbor spin correlation in the next section.

\section{Magnetization and Nearest-Neighbor Spin-Spin Correlation from Mean Field Theory}

\begin{figure*}
\begin{center}
\includegraphics[width=0.9\textwidth]{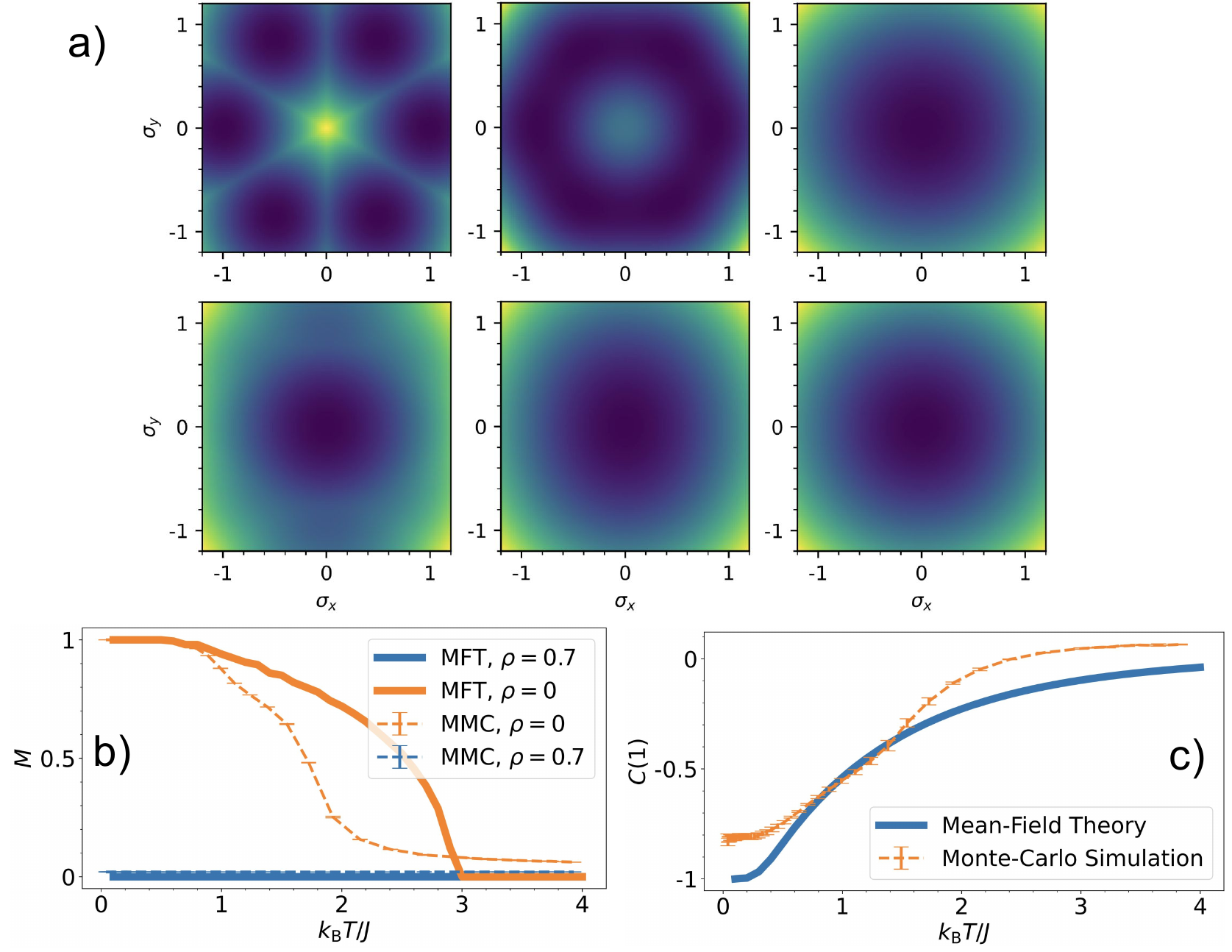}
\end{center}
\caption{
(a) Mean field free energies for the SF6C model, where $\sigma_x$ and $\sigma_y$ represent the $x$ and $y$ components of the net magnetization. 
The top row shows the unfrustrated model while the bottom row shows the model with $Q=+10$ ions and $\rho=\frac{7}{10}$. 
From left to right, the temperatures shown are $0.1 \frac{J}{\kb}$, $1.5 \frac{J}{\kb}$, and $5 \frac{J}{\kb}$. 
(b) Magnetization for the SF6C model with $Q=+5$ ions at $\rho=0$ and $\rho=0.7$ compared to the mean field result from a system with $Q=+10$ ions. Note that for $\rho=0.7$, the MFT curve is identically 0 for all temperatures while the MMC curve is slightly above 0 due to finite size effects but still constant within error bars.
(c) Nearest-neighbor spin-spin correlation function, $C(1)$, for the SF6C model with $Q=+5$ ions at $\rho=0.7$ and MFT with six-fold symmetry and $Q=+10$ ions at $\rho=0.7$. 
}
\label{fig:MFT}
\end{figure*}

To understand some of the novel behavior of our models, we developed a mean field theory (MFT) using a pair approximation, the details of which may be found in Appendix~\ref{sec:mft}. 
In the absence of ions, the unfrustrated 6-clock model exhibits a phase transition with temperature; the mean field free energy exhibits six degenerate minima at low $T$ and the system transitions to a single minimum at zero at high temperatures. 
When ions are added to create the SF6C model, and for a large enough ionic charge $Q$, there is a critical ionic density $\rho_c$ above which the only free energy minimum occurs with zero net magnetization independent of temperature.
That is, the MFT correctly predicts that for some ion densities, there is only a paramagnetic phase. 
This is demonstrated by the mean field free energies in in Fig.~\ref{fig:MFT}a, where the free energy for systems with high ionic concentrations exhibits only a minimum at zero even at very low temperatures. 
%

\begin{figure}
\begin{center}
\includegraphics[width=0.45\textwidth]{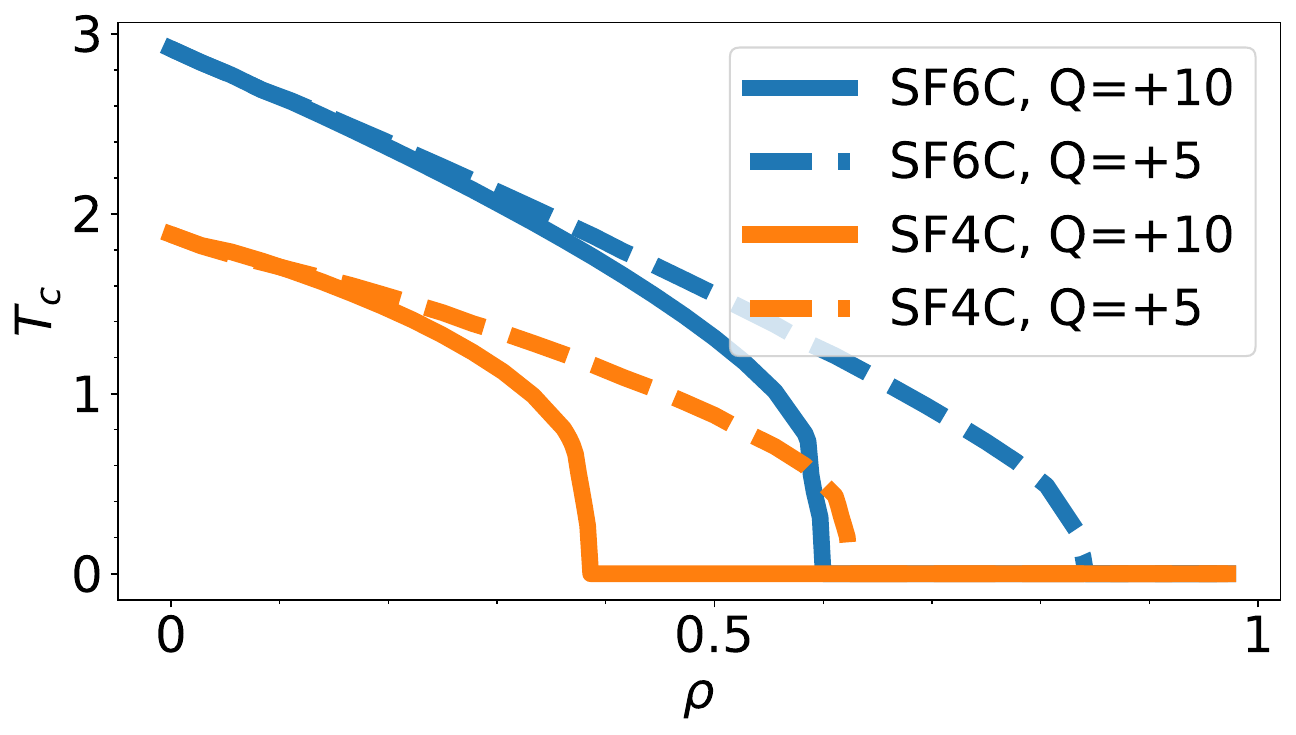}
\end{center}
\caption{
Estimated paramagnetic transition temperature, $T_c$, as a function of $\rho$ in the SF4C and SF6C models at $Q=+5$ and $Q=+10$ using mean-field theory.
}
\label{fig:mfttcvsrho}
\end{figure}

%
The destruction of order in the system upon addition of ions is also evidenced by the magnetization, Fig.~\ref{fig:MFT}b. 
We find that at low ionic concentrations, there is spontaneous symmetry breaking, whereby the magnetization becomes non-zero at low temperatures. 
However, at high ionic concentrations, this transition disappears, as we also observe in our Monte Carlo simulations. 
Note however that as is typical of mean field theories, our predictions are qualitative. 
To better match the Monte Carlo results with $Q=+5$ ions, we used $Q=+10$ in our MFT. 
Using this mean field approach, we compute the paramagnetic transition temperature for each model as a function of the ion concentration for both the SF4C and SF6C models with $Q=+5$ and $Q=+10$, Fig.~\ref{fig:mfttcvsrho}.
The critical temperature decreases with $\rho$ until a phase transition no longer occurs above a critical concentration, $\rho_c$. 
This critical concentration decreases with increasing $Q$, such that $\rho_c(Q=10)<\rho_c(Q=5)$ for both models. 
For the SF6C model, we find $\rho_c\simeq 0.600$ for $Q=+10$ and $0.842$ for $Q=+5$; for the SF4C model, we find $\rho_c\simeq 0.386$ and $0.625$ for $Q=+10$ and $+5$, respectively.
%

%
Our MFT is also able to predict the nearest-neighbor spin-spin (NNSS) correlation function, given by $C(1)=\langle \cos(\theta_i-\theta_j)\rangle,$ where the average is performed over nearest neighbor spin pairs $i$ and $j$. 
This measures the local, rather than global, alignment of spins. 
Remarkably, MFT provides qualitatively consistent values here as well. 
For systems with a net magnetization, we find a clear transition from a disordered phase with $C(1)\simeq 0$ to an ordered phase with $C(1)\simeq 1$. 
However, we also find that for systems with no net magnetization, there is still a clear order-disorder transition. 
Specifically, $C(1)\simeq -1$ in the ordered phase. 
This is representative of an ordered non-magnetic phase caused by spins anti-aligning with their neighbors due to the ions, which is also seen in our simulations. 
While the values given by MFT are not quantitatively consistent with the simulations, they do match rather well qualitatively. 
We show the curve of $C(1)$ from both MFT and a qualitatively similar Monte-Carlo result in Fig.~\ref{fig:MFT}c. 
Note that both the MFT and Monte Carlo systems do not have any net magnetization, but there is still a clear order-disorder transition at low temperatures. 

\section{Topological Phases}\label{sec:topphases}

We now turn to topological effects in the solute-frustrated $q$-clock models. 
Ions provide an interesting source for topological effects---in the high charge limit, where $KQ\gg J$, spins will point radially outward from an ion. 
The spins thus form a defect with a +1 topological charge, a vortex.
The interplay among these fixed topological defects, spin-spin ordering and its $q$-dependence (no toplogical transitions for $q<5$, topological transitions for $q>5$), and solute-solute interactions leads to non-trivial phase diagrams and topological behavior that differ in all four systems studied.
The SF4C model with +1 ions does not exhibit topological phases --- the 4-clock model itself does not exhibit topological defects and an associated BKT transition, and ions with $Q=+1$ do not create interactions strong enough to maintain fixed vortices. 
Therefore, we focus our discussion of topological phases on the other three systems studied, which all display topological effects. 

\subsection{BKT Critical Region}

\begin{figure}
\begin{center}
\includegraphics[width=0.45\textwidth]{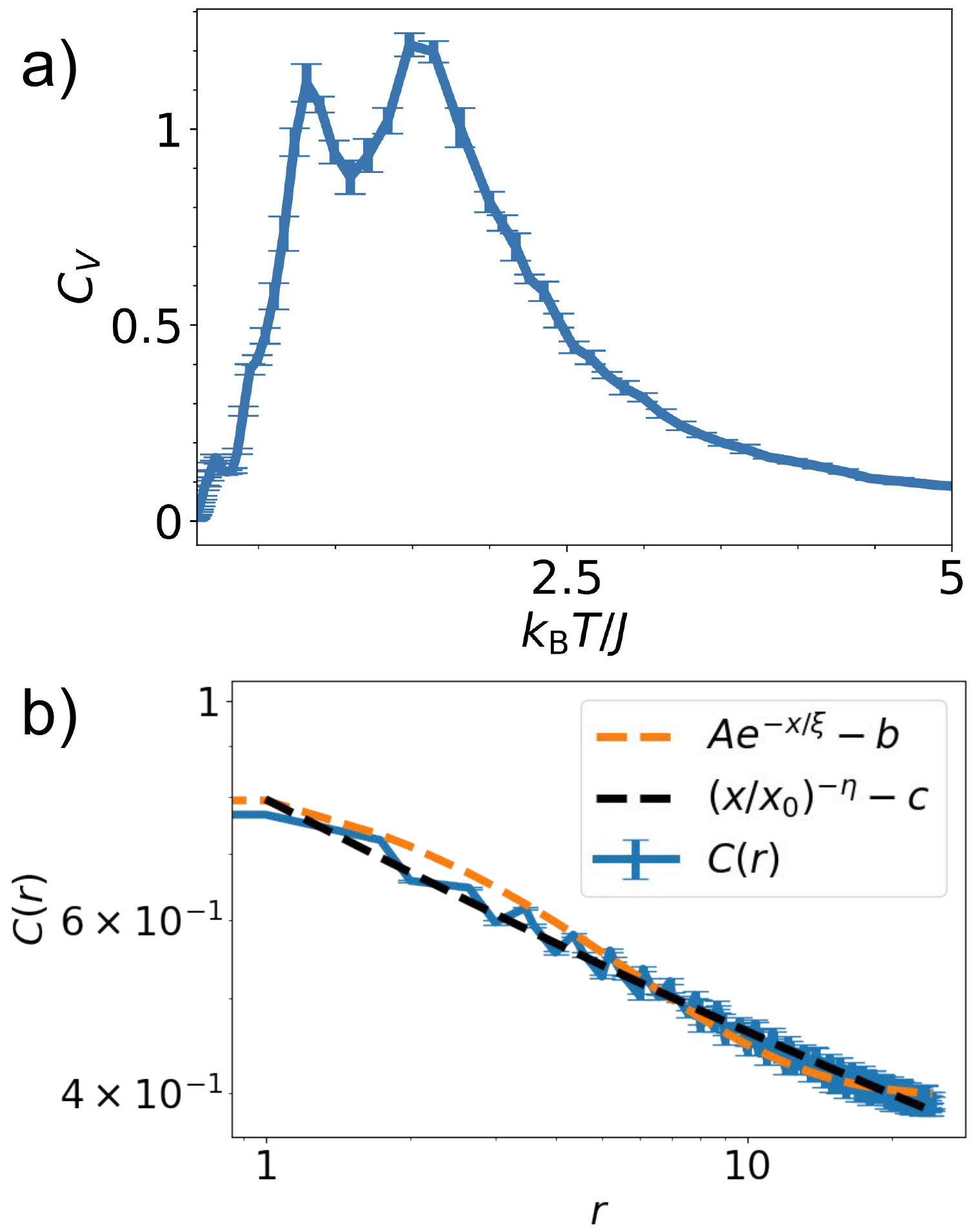}
\end{center}
\caption{
(a) Heat capacity of the SF6C model with $Q=+1$ ions and $\rho=\frac{1}{18}$. 
Note the dip in the heat capacity around $\kb T/J=1$. 
(b) Spin-spin correlation function in the SF6C model with $Q=+1$ ions and $\rho=\frac{1}{18}$ along with exponential and power law fits. 
}
\label{fig:bkt}
\end{figure}

The solute-free, unfrustrated $q$-clock model exhibits three phases for $q\geq5$ --- a low temperature broken-symmetry ferromagnetic phase, a high temperature paramagnetic phase, and a BKT critical region separating the low and high temperature phases. 
The BKT region is characterized by topological defects (vortices) and quasi-long-range order quantified by a spin-spin correlation function that decays algebraically. 
The nature of the higher temperature transition is rather debated, though many argue that it is of BKT-type~\cite{jose1977,elitzur1979,tobochnik1982}. 
The BKT critical region is preserved in the SF6C models for low ion concentrations. 
To quantify the existence of this phase, we focus on the heat capacity,
\begin{equation}
C_V=\frac{\expval{E^2}-\expval{E}^2}{\kb T^2},
\end{equation}
where $E$ is the energy,
and the spin-spin correlation function
\begin{equation}
C(r)=\langle\mathbf{S}_i(\mathbf{0})\cdot\mathbf{S}_j(\mathbf{r})\rangle=\left\langle \cos(\theta_i(\mathbf{r}_i)-\theta_j(\mathbf{r}_j))\right\rangle,
\end{equation}
where $\theta_i(\mathbf{r}_i)$ represents a spin at position $\mathbf{r}_i$; note that we average over all pairs $(i,j)$ such that $\|\mathbf{r}_i-\mathbf{r}_j\|=\|\mathbf{r}_{ij}\|=r$. 
In Fig.~\ref{fig:bkt}a, we see a dip in the heat capacity of the SF6C model with $Q=+1$ ions at $\rho=\frac{1}{18}$ centered around $\kb T/J=1$, with two strong peaks on either side. 
The double-peaked structure of $C_V$ indicates that there are two distinct phase transitions which take place in the model, similar to the two transitions that occur in the unfrustrated 6-clock model.
This provides evidence that the three phases in the 6-clock model have analogues when ions are added, including the BKT critical region.
We further demonstrate this by analyzing the decay of the spin-spin correlation function, which can distinguish between different phases with the same magnetization~\cite{bramwell1993,bramwell1994,archambault1997}. 
Specifically, for a paramagnetic system, the lack of long-range correlations leads to an exponentially decaying correlation function, $C(r)\sim e^{-\frac{r}{\xi}}$.
For a ferromagnetic system with infinite-range order, $C(r)$ is approximately constant~\cite{landaubinder2014}. 
However, BKT-like phases exhibit quasi-long range order, such that the correlation function decays algebraically, $C(r)\sim \left(\frac{r}{r_0}\right)^{-\eta}$~\cite{berezinskii1971,kosterlitz1973,jose1977}. 
We find that the BKT-like phases in the SF6C models indeed exhibit algebraic decay, Fig.~\ref{fig:bkt}b. 
We find similar results for the SF6C model with $Q=+5$ and $\rho=\frac{1}{3}$ at low temperatures, where the system acts identically to an unfrustrated 6-clock model on a honeycomb lattice, which also has the BKT critical region. 
A more stringent quantification of BKT phase transitions involves the computation and analysis of the helicity modulus~\cite{nelson1977}, but we reserve this for future work.
%

\subsection{Ionic Cell Walls}\label{subsec:cells}

\begin{figure}
\begin{center}
\includegraphics[width=0.45\textwidth]{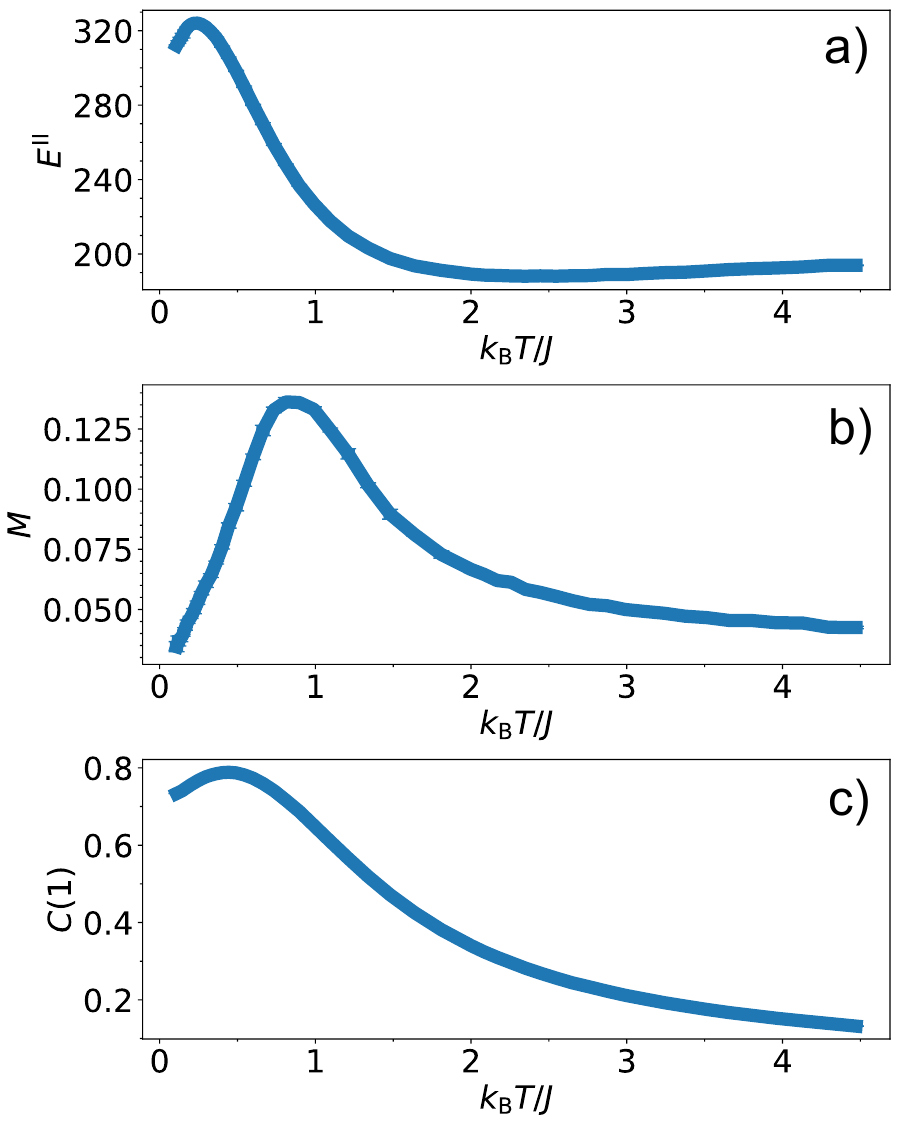}
\end{center}
\caption{
(a) Ion-ion energy, (b) magnetization, and (c) nearest-neighbor spin-spin correlation function in the SF6C model with $Q=+1$ ions at $\rho=\frac{1}{4}$. 
}
\label{fig:ionwallops}
\end{figure}

%
The introduction of solute frustration also creates new phases not found in the unfrustrated clock models,
one of which we refer to as ionic cell wall phases, Fig.~\ref{fig:sf6c_q1_phases} and~\ref{fig:cellwallmelting}a.
In the SF6C model with $Q=+1$, ion-ion repulsions are overwhelmed by spin-spin and spin-ion interactions, such that the ion assemble into long 1D chains.
At low concentrations, the ions assemble into a single long strand, while at high concentrations the ionic chains connect to form cells of spins surrounded completely by ionic cell walls. 
At even higher concentrations, lone ions are found at the center of these cells. 

\subsubsection{Energetics of Cell Wall Formation}

\begin{figure}
\begin{center}
\includegraphics[width=0.45\textwidth]{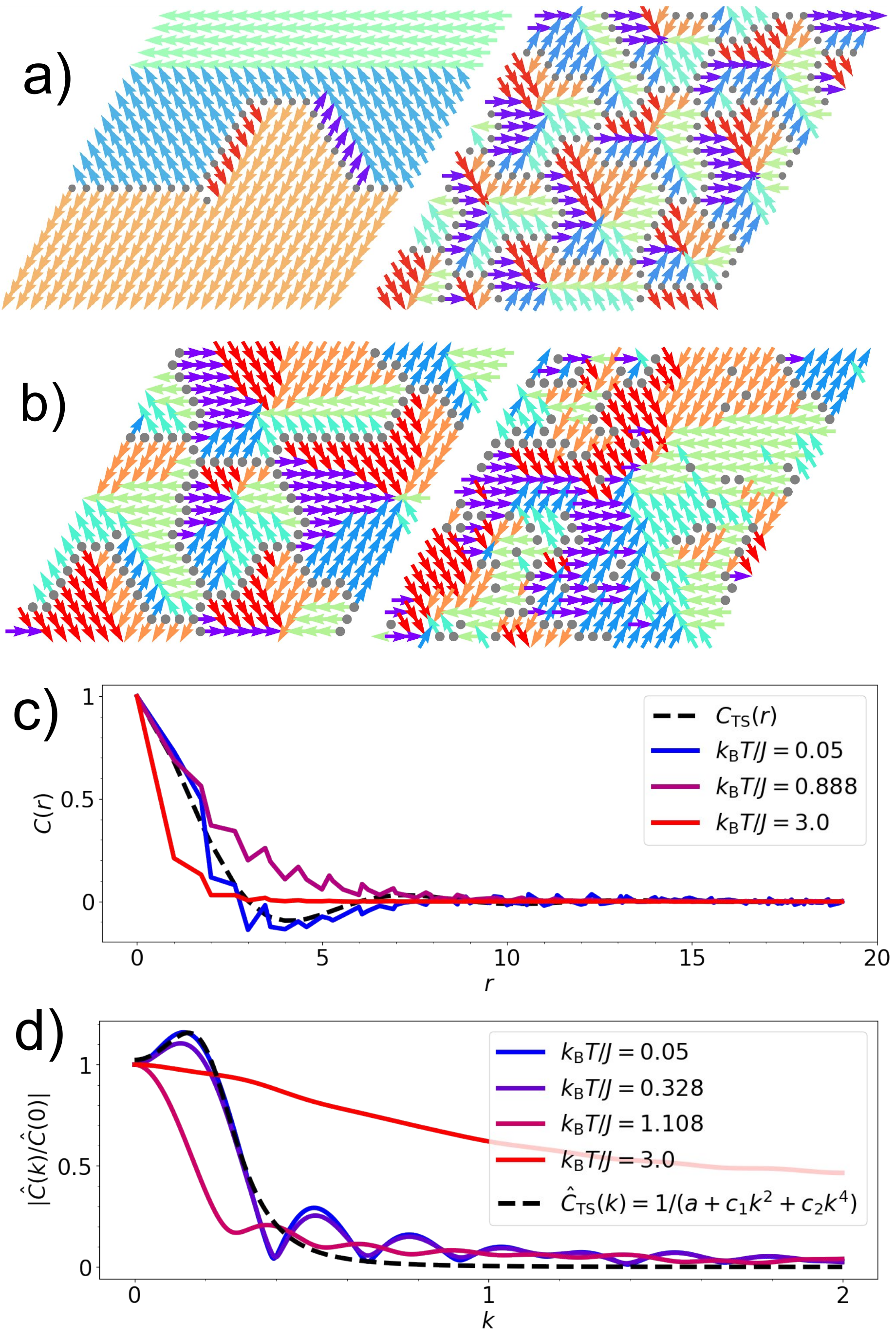}
\end{center}
\caption{
(a) Cell walls in the SF6C model with +1 ions at $\rho=\frac{1}{18}$ (left) and $\rho=\frac{1}{4}$ (right) at $T=0.04 \frac{J}{\kb}$. 
(b) Melting of the SF6C model with +1 ions at $\rho=\frac{1}{6}$ between (left) $T=0.04\frac{J}{\kb}$ and (right) $T=0.81\frac{J}{\kb}$. 
(c) Spin-spin correlation function of the SF6C model with $Q=+1$ ions at $\rho=\frac{1}{4}$ as a function of temperature. 
(d) Structure factor of the SF6C model with $Q=+1$ ions at $\rho=\frac{1}{18}$ as a function of temperature. 
The structure factors have been normalized for convenience. 
The dashed line fits the lowest temperature structure factor to the Teubner--Strey model~\cite{teubner1987}.
}
\label{fig:cellwallmelting}
\end{figure}

We can quantify the transition to cell walls through the individual spin-spin, spin-ion, and ion-ion components to the energy.
We find the most useful to be the ion-ion energy,
\begin{equation}
E^{\mathrm{II}}=\left\langle LQ^2\sum_{\langle i,j\rangle}t_it_j\right\rangle.
\end{equation}
During the transition to the cell wall phase, $E^{\mathrm{II}}$ sharply increases as ionic chains form, Fig.~\ref{fig:ionwallops}a.
At high temperatures, entropy drives the melting of these chains, Fig.~\ref{fig:cellwallmelting}b, and the ion-ion energy strongly decreases.
As the cell walls melt, the size of the magnetic domains increases, leading to a peak in the magnetization, Fig.~\ref{fig:ionwallops}b. 
However, the domains are finite and their orientation random, such that the magnetization vanishes in the thermodynamic limit. 
The local, short-range structure of these domains is further evidenced by the NNSS correlation, $C(1)$, which remains finite as $T\rightarrow0$, indicating that spins remain locally aligned, Fig.~\ref{fig:ionwallops}c.

%
At low ionic concentration, there is a temperature-dependent transition between smooth and rough cell walls, with the latter often containing free ions in each cell.
At low temperature, a surface tension stabilizes smooth cell wall structures, limiting deformation of the ionic cell walls. 
At high temperatures, entropy-driven fluctuations result in highly deformed or rough ionic cell walls. 
As the ionic concentration is increased, these fluctuations disrupt the ionic walls to the extent that the system becomes indistinguishable from the magnetic domain phase. 
At the highest concentrations where ionic cell walls form, the cell walls can become several ions thick, resulting in the thick cell wall phase.  
Note that because the walls are one-dimensional, this is a crossover and not a roughening transition~\cite{ChuiWeeks1976,Burton1951,Weeks1980,Fisher1984}.

\subsubsection{Structure of Cell Walls}

We quantify the structure of the ionic cell wall phases through the spin-spin correlation function, $C(r)$, Fig.~\ref{fig:cellwallmelting}c. 
At high temperatures, $C(r)$ decays exponentially, as expected for a paramagnetic or magnetic domain phase. 
As the temperature is lowered, additional structure appears, and $C(r)$ exhibits a minimum at negative values before decaying to zero. 
The appearance of this minimum in $C(r)$ is indicative of a second length scale appearing in the system, here due to the formation of ionic walls. 
This second length scale also appears in the structure factor,
\begin{equation}
\hat{C}(k)=\sum_r C(r)e^{-ikr},
\end{equation}
where $k$ is a wavevector, Fig.~\ref{fig:cellwallmelting}d.
At high temperature, $\hat{C}(k)$ exhibits a single broad peak centered at $k=0$, as expected for a paramagnetic phase. 
At low temperatures, a peak at finite $k$ appears, consistent with the second length scale discussed above. 
The appearance of a second length scale in $\hat{C}(k)$ due to microphase separation can be described by the Teubner-Strey model~\cite{teubner1987}, which was originally introduced to describe scattering from bicontinuous microemulsions~\cite{teubner1987} and can be derived from the charge-frustrated Ising model~\cite{woo1995,deem1994a}.
Within this perspective, the cell wall phase is an analogue of a microemulsion; the solute walls play the role of a surfactant film and the magnetic domains are akin to oil and water domains, as in lattice models of microemulsions~\cite{widom1986,levin1990,levin1992}.
Therefore, we can fit the structure factor to the Teubner-Strey form, 
\begin{equation}
\hat{C}_{\rm TS}(k)=\frac{1}{a+c_1k^2+c_2k^4},
\end{equation}
where $a$, $c_1$, and $c_2$ are fit parameters. 
Inverting $\hat{C}_{\rm TS}(k)$ yield the Teubner-Strey estimate of the spin-spin correlation function,
\begin{equation}
C_{\rm TS}(r)=\frac{d}{2\pi r}e^{-r/\xi}\sin\left(\frac{2\pi r}{d}\right),
\end{equation}
where the correlation length for the magnetic domains is
\begin{equation}
\xi=\frac{1}{\sqrt{\sqrt{\frac{a}{4c_2}}+\frac{c_1}{4c_2}}}
\end{equation}
and domain repeat distance
\begin{equation}
d=\frac{2\pi}{\sqrt{\sqrt{\frac{a}{4c_2}}-\frac{c_1}{4c_2}}},
\end{equation} 
such that the negative minimum we observe in $C(r)$ is the real space signature of microemulsion-like domains~\cite{teubner1987,woo1995}.
In both real and reciprocal space, we find good agreement between the Teubner-Strey model and the simulated correlation functions, particularly the negative minimum in real space and the peak at finite $k$ in reciprocal space, Fig.~\ref{fig:cellwallmelting}.
Note that in reciprocal space, we performed the curve fits for low $k$, to avoid errors associated with ringing at higher $k$ due to finite size effects.
%

\begin{figure}
\begin{center}
\includegraphics[width=0.45\textwidth]{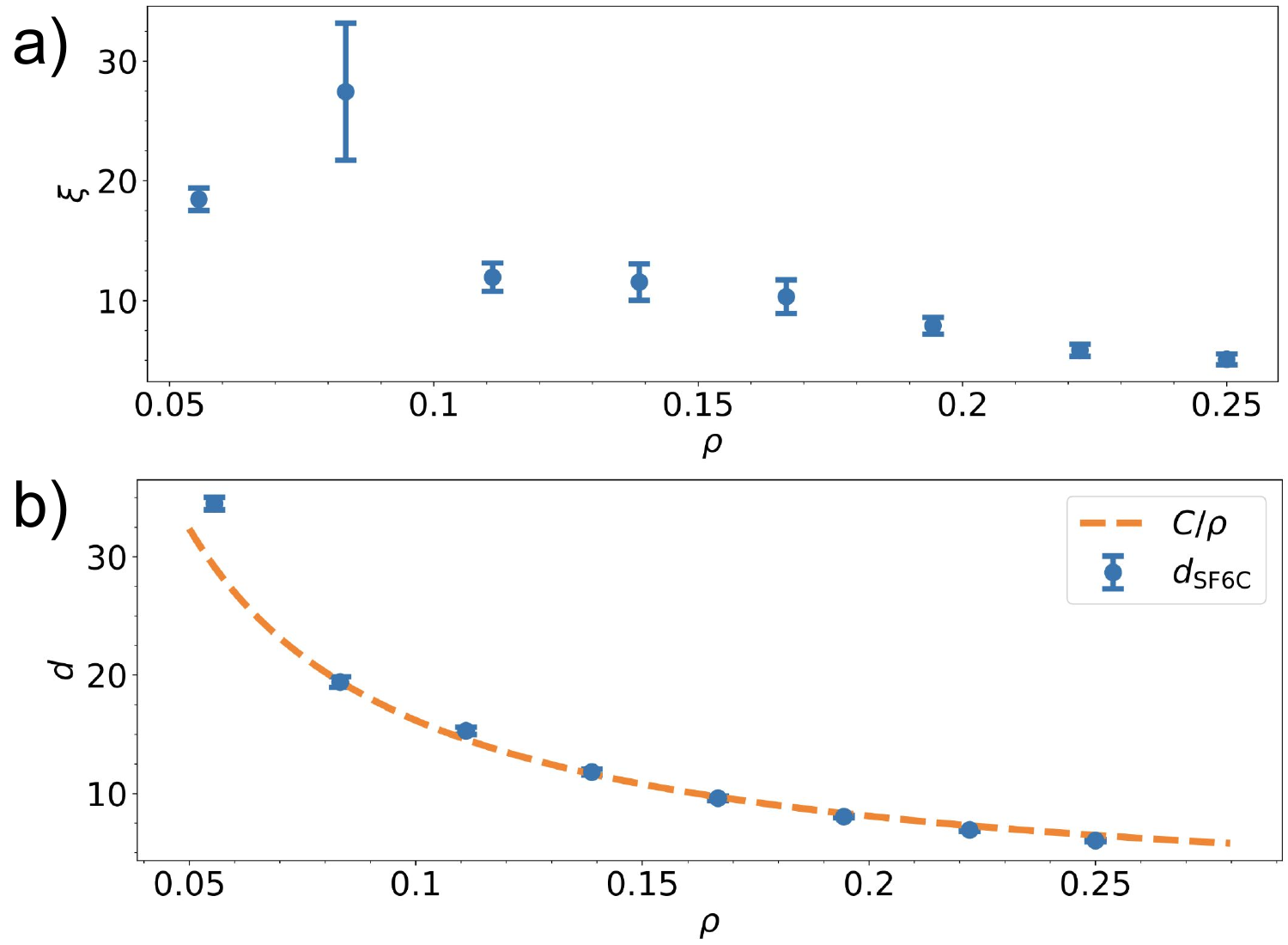}
\end{center}
\caption{
(a) Fitted Teubner-Strey magnetic domain correlation lengths, $\xi$, against the ionic concentration $\rho$ in the SF6C model with $Q=+1$ ions at $\frac{\kb T}{J}=0.05$. (b) Fitted Teubner-Strey domain repeat distances, $d$, against the ionic concentration $\rho$ in the SF6C model with $Q=+1$ ions at $\frac{\kb T}{J}=0.05$ compared with $\frac{1}{\rho}$ curve.
}
\label{fig:teubnerstreyfitting}
\end{figure}

%
We calculated $\xi$ and $d$ over a range of concentrations, shown in Fig.~\ref{fig:teubnerstreyfitting}.
We find that there is a general trend for $\xi$ to decrease with increasing $\rho,$ though with fairly high uncertainty, likely due to finite size effects having a strong impact on the size of magnetic domains.
On the other hand, the trend for $d$ is very clear, with the results indicating that $d\sim \frac{1}{\rho}.$
This is easy to show theoretically via a simple scaling argument --- consider tiling a fixed plane of size $\ell^2$ by $N_{d}$ domains bounded by ions, with $d$ being the length scale of these domains.
The area of each domain is proportional to $d^2,$ so we find that 
\begin{equation}
N_d d^2\sim \ell^2.
\label{eq:areaproportionality}
\end{equation}
The perimeter of each domain is proportional to $d$, so the total boundary length of the domains is proportional to $N_d d$.
Since these boundaries are formed by ions, we know that the number of ions, $\rho\ell^2,$ is proportional to the total boundary length, so \begin{equation}
N_d d\sim \rho\ell^2.
\label{eq:perimeterproportionality}
\end{equation}
Comparing Eqs.~\ref{eq:areaproportionality} and \ref{eq:perimeterproportionality}, we see that $d\sim \frac{1}{\rho},$ as we observe in the data, Fig.~\ref{fig:teubnerstreyfitting}

\subsubsection{Topology Within Cells}

\begin{figure*}[t]
\begin{center}
\includegraphics[width=0.9\textwidth]{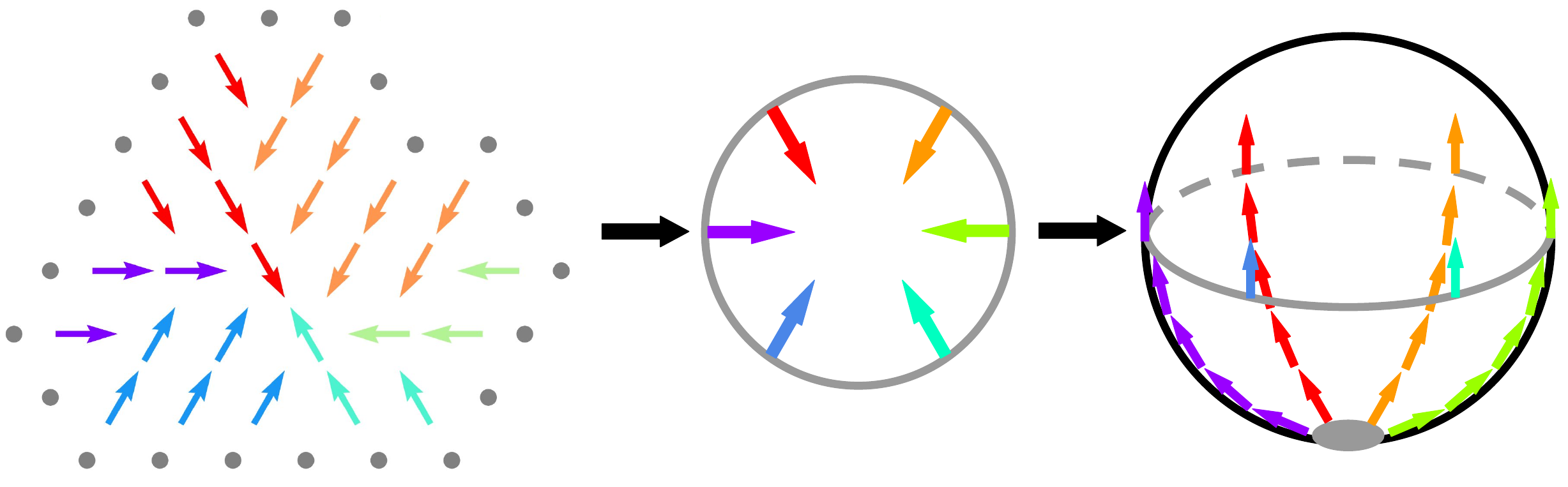}
\end{center}
\caption{
Transformation of a +1 vortex inside a cell wall in the SF6C model to a sphere with a positive vortex at one of its poles. 
}
\label{fig:topologyofwalls}
\end{figure*}

%
Unlike traditional microemulsions and microphase separation found in the charge-frustrated Ising model, the cell wall phases in the SF$q$C models have non-trivial topology.
As illustrated by the snapshots in Fig.~\ref{fig:cellwallmelting}a and~b, and Fig.~\ref{fig:topologyofwalls}, each cell or domain contains a positive vortex near its center. 
As a result, each cell contains an unbound and seemingly unbalanced positive vortex even near zero temperature. 
The existence and stability of these vortices, as well as why they exist near the center of a cell, can be rationalized from the topology of the system. 
The ionic walls effectively shield one cell from the others, such that we can consider each cell as an isolated system.
Moreover, the ionic walls enforce a specific boundary condition where spins immediately adjacent to the boundary will point radially inwards due to the spin-ion interactions; a circular arrangement of ions to form a wall forces the spins to point toward the center of the cell's interior. 
To understand how the remaining spins in the cell orient, we can work in analogy to the method of images used in classical electrostatics~\cite{Zangwill}. 
Consider a sphere with a +1 topological charge located at its south pole, such that spins point radially outwards from the pole, Fig.~\ref{fig:topologyofwalls}.
Then, on the equator, as viewed from the north pole, spins will point radially inwards. 
This is the same boundary condition as in the cell walls --- continuous deformation leads to a circular cell wall, topologically equivalent to the physical cell wall. 
Therefore, the cell walls in the SF$q$C models are akin to the northern hemisphere of a sphere with a +1 charge at its south pole. 
By the Poincar\'{e}-Hopf Theorem on a sphere with genus zero, the sum of the topological charges must equal 2; this special case is known as the Hairy Ball Theorem~\cite{mermin1979}. 
Because we have a +1 charge at the south pole and no other charges in the system, the northern hemisphere must also have a +1 charge, such that each cell must contain a +1 topological charge within it. 
Moreover, because like topological charges repel, the charge in the northern hemisphere is pushed toward the north pole, consistent with an unbound vortex being near the center of each cell.
To maintain global topological neutrality, the ionic walls carry the neutralizing negative charge, though this may not be immediately obvious when the cells are large. 
Boundary condition-induced defects of this type appear often in confined nematic liquid crystals, where anchoring conditions at surfaces constrain the total charge of the interior defect~\cite{mermin1979}, leading to topological defect-driven self-assembly of colloidal particles into two-dimensional crystals~\cite{musevic2006}.

\subsection{Topological Dipoles and Defect-Driven Crystallization}

\begin{figure}
\begin{center}
\includegraphics[width=0.45\textwidth]{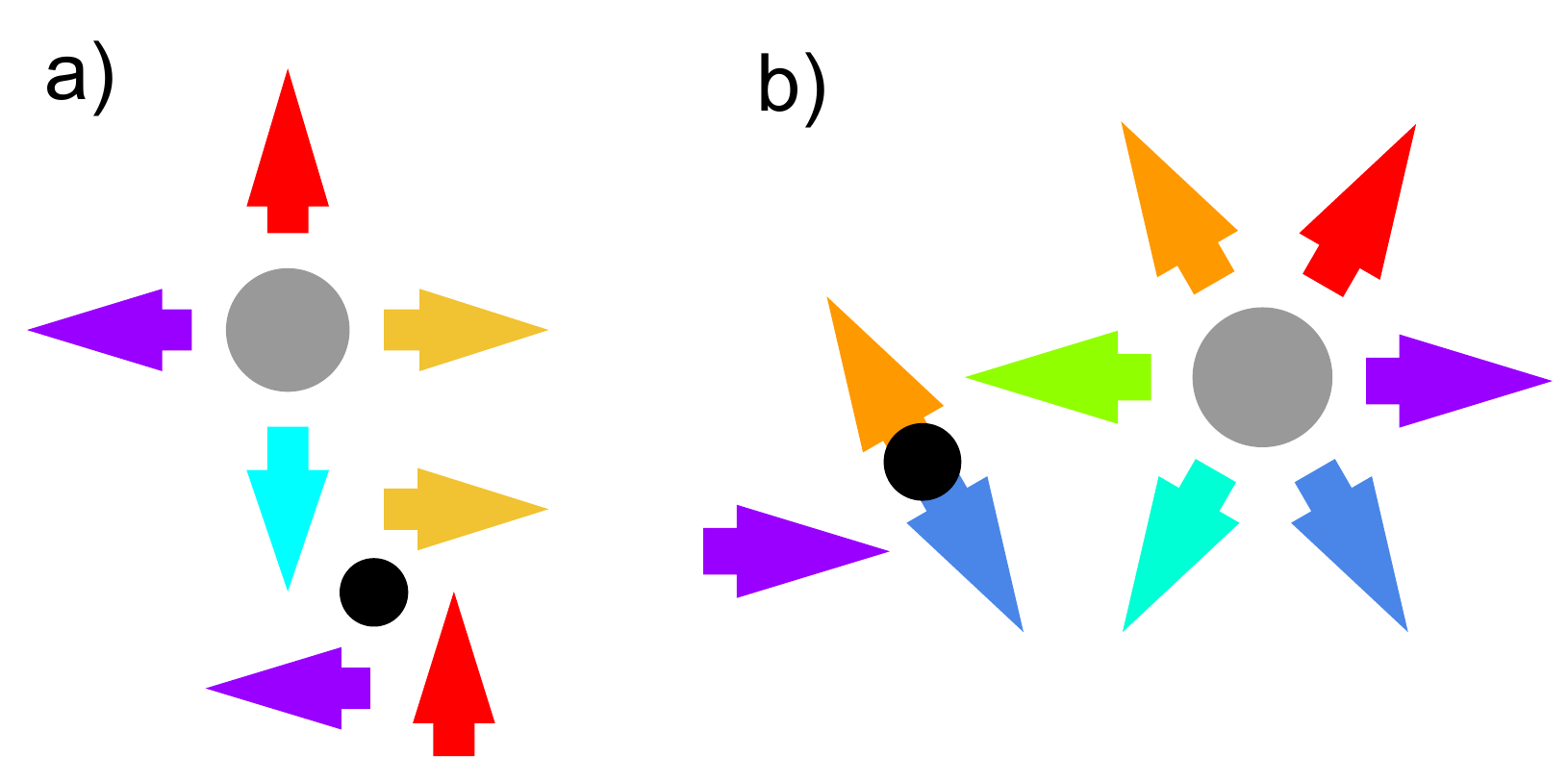}
\end{center}
\caption{
Topological dipoles in the (left) SF4C and (right) SF6C models. 
The black circle shows the center of the negative topological charge, while the grey ion is the center of the positive charge.
}
\label{fig:topologicaldipole}
\end{figure}

When ion charges are large, $Q=5$, the surrounding spins point radially outward, creating fixed ion-centered vortices in the SF$q$C model. 
Unlike the XY model, there is no mechanism for vortex-antivortex annihilation in the SF$q$C models, such that each ion creates a fixed lower bound on the number of vortices in the system. 
To satisfy neutrality, each vortex induces the formation of an antivortex (topological charge $-1$), which is Coulombically attracted to the vortex to form a bound vortex-antivortex pair. 
Each stable bound vortex-antivortex pair creates a topological dipole, Fig.~\ref{fig:topologicaldipole}. 
Analogous topological dipoles are known to form in the frustrated XY model, in which the bond energy, $-J\cos(\Delta\theta-A_{\mu})$, includes a contribution from a gauge field, $A_\mu$, representing a magnetic vector potential that forces the creation of topological dipoles (bound vortex-antivortex pairs)~\cite{teitel1983}.
However, the gauge field in the frustrated XY model and the resulting topological dipoles are fixed. 
In contrast, because the dipoles are attached to mobile ionic sites, the topological dipoles are free to translate and rotate, enabling them to assemble into unique phases. 
%

%
Both the SF4C and SF6C models exhibit topological defect-induced crystallization, and the specific crystalline phase that forms depends on $q$ and the symmetry of the lattice. 
To understand what phases may exist in each system, we can simplify our treatment of the model by
considering a system with equal amounts of positive and negative topological charges and neglecting all other interactions.
The system exists on a lattice, which may not be the physical lattice used in our simulations, with all sites occupied.
Then, the system can be described by sites with $n_i=\pm 1$ and an effective Hamiltonian
\begin{equation}
\mathcal{H}_{\mathrm{eff}}=-U_d\sum_{i,j}n_in_j\ln\left(\frac{|\mathbf{x}_i-\mathbf{x}_j|}{L_d}\right),
\end{equation}
where $U_d$ and $L_d$ are constants and $\mathbf{x}_i$ is the position of site $i$. 
In this limit of densely-packed defects, we approximate the Hamiltonian with only nearest-neighbor interactions,
\begin{equation}
\mathcal{H}_d=-J_d\sum_{\langle i,j\rangle}n_in_j,
\end{equation}
where we have introduced a new coupling constant $J_d<0$.
This approximate Hamiltonian is that for an antiferromagnetic (AFM) Ising model with no external magnetic field.
Within this approximation, it follows that, like the AFM Ising model, the SF4C model on a square lattice has a finite temperature phase transition involving an ordered state with up-down spin pairs, analogous to topological dipoles. 
As a result, the SF4C model should exhibit crystallization of the topological dipoles. 
In contrast, the AFM Ising model is frustrated on a hexagonal lattice, such that a similar topological defect-driven crystal cannot form in the SF6C model; we show below that a qualitatively different topological defect-driven crystallization can occur in this model.
We can loosen our restriction on site occupancy to include empty sites, such that $n_i=0,\pm1$, resulting in a spin-1 Ising model. 
For notational consistency with the Ising model, we rewrite $n_i$ and $s_i$, the spin orientation of site $i$.
We work in the grand canonical ensemble to enforce the correct number of charges, and we define $N_0=\sum_{i}(1-s_i^2)$ as the number of unoccupied (spin-0) sites and $N$ as the total number of sites, such that 
$\frac{N-N_0}{2}$ is the number of positive topological charges or ions in the SF$q$C models.
Keeping $N_0$ constant, we have 
\begin{equation}
\mathcal{H}_{\mathrm{eff}}-\mu N_0=-U_d\sum_{i,j}s_is_j\ln\left(\frac{|\mathbf{x}_i-\mathbf{x}_j|}{L_d}\right)-\mu N_0,
\end{equation}
where $\mu$ is the chemical potential. 
If we once again make the approximation of only nearest neighbor interactions, this reduces to
\begin{equation}
\mathcal{H}_d-\mu N_0=-J_d\sum_{\langle i,j\rangle}s_is_j-\mu \sum_{i}(1-s_i^2).
\end{equation}
Noting that $\sum_i 1=N$, we can set $E_0=-\mu N$ and define $D=-\mu$, such that the Hamiltonian becomes 
\begin{equation}
\mathcal{H}_{\mathrm{AFM}}=\mathcal{H}_d-\mu N_0=E_0-J_d\sum_{\langle i,j\rangle}s_is_j-D\sum_{i}s_i^2,
\end{equation}
the Hamiltonian for a spin-1 AFM Ising model with anisotropy coefficient $D$, or the Blume-Capel model~\cite{blume1966,capel1966}. 
The anisotropy coefficient, $D$, corresponds to an external field that acts equally on both up and down spins while having no interaction with spin zero sites.
In the analogy to the SF$q$C models, $D$ is related to the chemical potential that controls the ion concentration. 
We can use this Blume-Capel model to further build expectations regarding the topological defect-driven phase behavior of the SF$q$C models. 
At zero temperature, the Blume-Capel models shows a transition at $D=-2$ between a state with only $s_i=0$ for $D<-2$ and a state with only $s_i=\pm1$ for $D>-2$. 
At finite temperature, the Blume-Capel model exhibits phase separation, suggested that the SF$q$C models can exhibit phase separation between regions of crystallized topological dipoles and regions without ions. 
Moreover, the Blume-Capel model exhibits three zero-temperature phases on a hexagonal lattice that suggest similar phases in the SF6C model.
At very low $D$, there will be no ions and the phase will be that of the 6-clock model.
At very high $D$, the SF6C model will exhibit a frustrated ground state analogous to the normal AFM Ising model. 
But, at moderate values of $D$, the Blume-Capel model exhibits a fully ordered non-degenerate phase consisting of alternating ``sheets" of up, down, and spin-0 states.

\subsubsection{Topological Defect-Driven Crystallization in the SF4C Model}

\begin{figure}
\begin{center}
\includegraphics[width=0.45\textwidth]{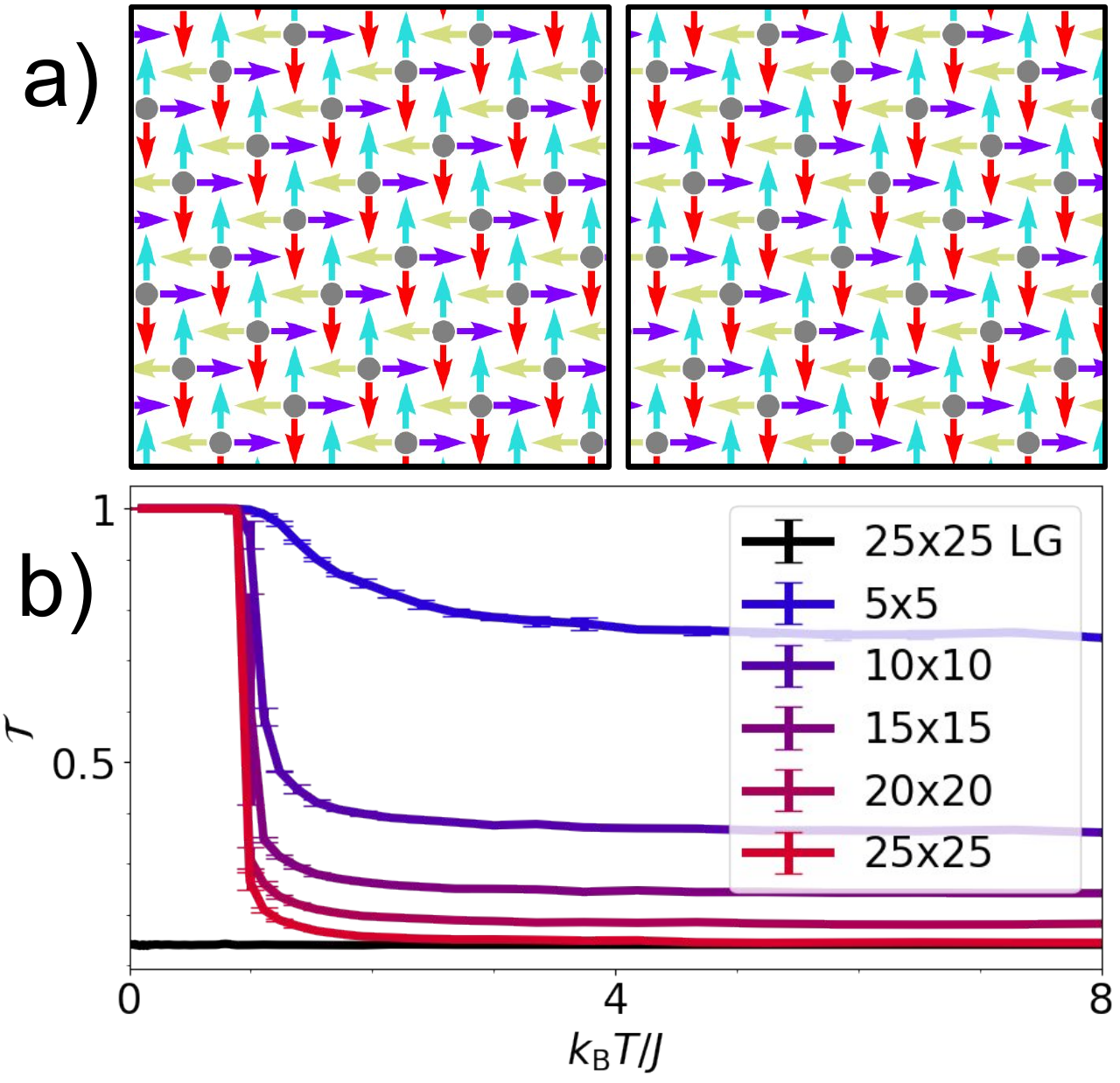}
\end{center}
\caption{
(a) Topological crystals in the SF4C model with $Q=+5$ and $\rho=\frac{1}{5}$ in the (left) left-handed form and (right) right-handed form. 
(b) Topological crystallization order parameter $\mathcal{T}$ in the SF4C model with $Q=+5$ and $\rho=\frac{1}{5}$ over various lattice sizes along with a lattice gas for comparison. 
}
\label{fig:topocrystals}
\end{figure}

The SF4C model with $Q=5$ exhibits topological crystallization at $\rho=\frac{1}{5}$.
The topological crystal lattice is not the physical lattice but instead a rotated and scaled square lattice.
The resulting topological crystal breaks mirror symmetry and can be either right- or left-handed, Fig.~\ref{fig:topocrystals}a.
In the center of each cell of the crystal, there is a positive topological charge consisting of spins wrapping around it. 
We choose the right-handed crystal such that the spins wrap counterclockwise around the charge --- the curl of the spins points out of the page --- and vice versa for the left-handed crystal. 
In both cases, the winding number remains +1 because topological charges act like electric charges and do not change sign when mirrored.
Because the underlying Hamiltonian is achiral, the formation of topological crystals with a specific handedness is an example of spontaneous mirror symmetry breaking, similar to chiral ordering in skyrmion crystals~\cite{muhlbauer2009}. 
Moreover, the topological crystal may also be viewed as an annealed analogue of the vortex superlattices that form in frustrated XY models~\cite{teitel1983}, except that the defects here are dynamic degrees of freedom tied to the mobile solute ions.
%

%
In this crystalline form, each ion is the center of a positive topological charge. 
Between two ions, there is a negative topological charge, and, in the center of four ions, there is another positive topological charge. 
This ensures total neutrality of the charges. 
To quantify topological crystallization, we define two order parameters, one for each handedness,
\begin{equation}
\mathcal{T}_{L}=\frac{1}{N}\sum_{i}t_ie^{\frac{2\pi i}{5}(2\mathbf{\hat{x}}+\mathbf{\hat{y}})\cdot\mathbf{r}_i},
\end{equation}
and 
\begin{equation}
\mathcal{T}_{R}=\frac{1}{N}\sum_{i}t_ie^{\frac{2\pi i}{5}(-2\mathbf{\hat{x}}+\mathbf{\hat{y}})\cdot\mathbf{r}_i},
\end{equation}
where $\mathbf{r}_i$ is the position of site $i$. 
In the right-handed ordered form, $\left|\mathcal{T}_R\right|>0$ while $\left|\mathcal{T}_L\right|=0$, and vice versa in the left-handed ordered form. 
Both are zero in the disordered phase. 
To generalize, we define a unified order parameter that accounts for both forms, 
\begin{equation}
\mathcal{T}=\left|\mathcal{T}_L\right|+\left|\mathcal{T}_R\right|.
\end{equation}
In the ordered phase, $\mathcal{T}>0$, while in the disordered phase, $\mathcal{T}=0$. 
The SF4C model with $Q=5$ exhibits a phase transition near $T\approx 0.9J/\kb$, as evidenced by the transition in $\mathcal{T}(T)$, Fig.~\ref{fig:topocrystals}b. 
Such a transition is absent in a pure lattice gas, also shown in Fig.~\ref{fig:topocrystals}b, indicating that this transition is mediated by spins and is topological. 
The transition in $\mathcal{T}(T)$ sharpens as the system size is increased, suggesting that the phase transition is first order, but we reserve a thorough finite-size analysis for future work~\cite{challa1986,leekosterlitz1990}.

\subsubsection{Topological Defect-Driven Crystallization in the SF6C Model}

\begin{figure}
\begin{center}
\includegraphics[width=0.45\textwidth]{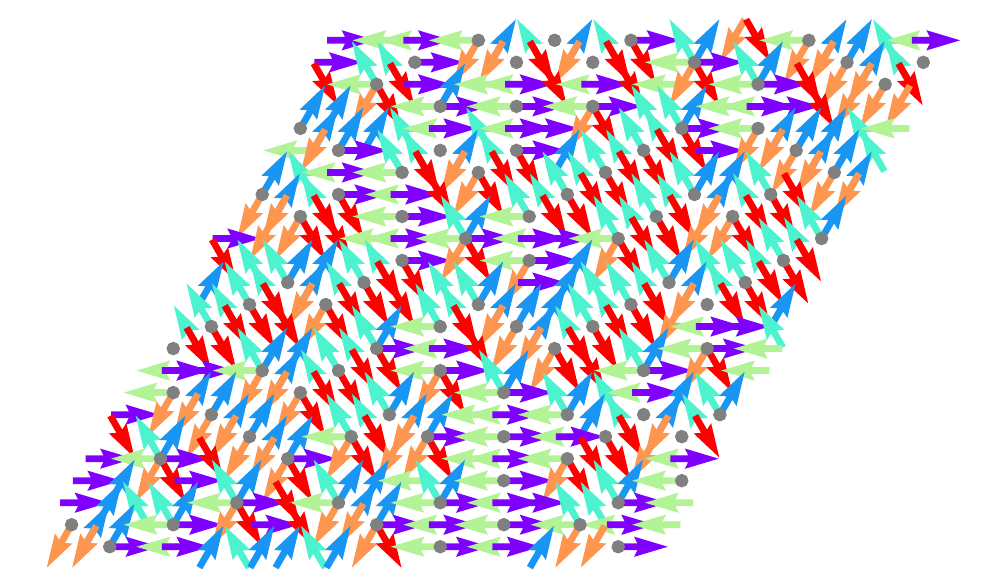}
\end{center}
\caption{
Low-temperature configuration of the SF6C model with $Q=+5$ ions and $\rho=\frac{1}{6}$. 
}
\label{fig:sf6c016666}
\end{figure}

The SF6C model with $Q=5$ cannot topological crystallize into a structure analogous to that found for the SF4C model due to the frustration induced by the hexagonal lattice; a spin-$\frac{1}{2}$ AFM Ising model also cannot crystallize on a hexagonal lattice. 
Instead, we observe long chains of ions held together by negative topological charges, but no extended or global two-dimensional structure forms, Fig.~\ref{fig:sf6c016666}.
This is analogous to the hexagonal AFM Ising model, which forms string-like configurations in the ground state but no long-range structure due to frustration.
%

\begin{figure}
\begin{center}
\includegraphics[width=0.45\textwidth]{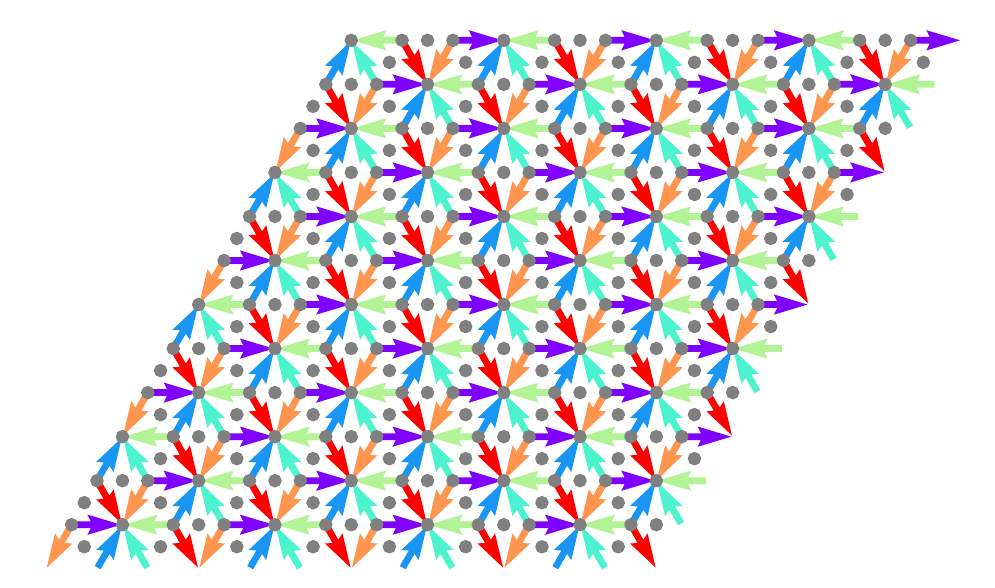}
\end{center}
\caption{
Low-temperature configuration of the SF6C model with either $Q=+1$ or $Q=+5$ ions at $\rho=\frac{1}{2}$. 
}
\label{fig:sf6c50000}
\end{figure}

%
The SF6C model can undergo topological crystallization if the system relieves frustration by forming a crystal that is not hexagonal.
This is achieved at $\rho=\frac{1}{2}$, when the system crystallizes into a regular hexagonal tiling with a single ion in the center, Fig.~\ref{fig:sf6c50000}.
The spins orient toward the central ion in a flower petal-like arrangement, and this arrangement may be thought of as a limit case of the cell walls. 
In this case, the central ion and the corners of each hexagon correspond to positive topological defects (vortices). 
The sides of the hexagon act as negative topological defects (antivortices) to balance the overall topological charge. 
The structure that results is a deltoidal trihexagonal tiling of the plane, which allows for an unfrustrated ground state of the antiferromagnetic Ising model. 
%

\begin{figure}
\begin{center}
\includegraphics[width=0.45\textwidth]{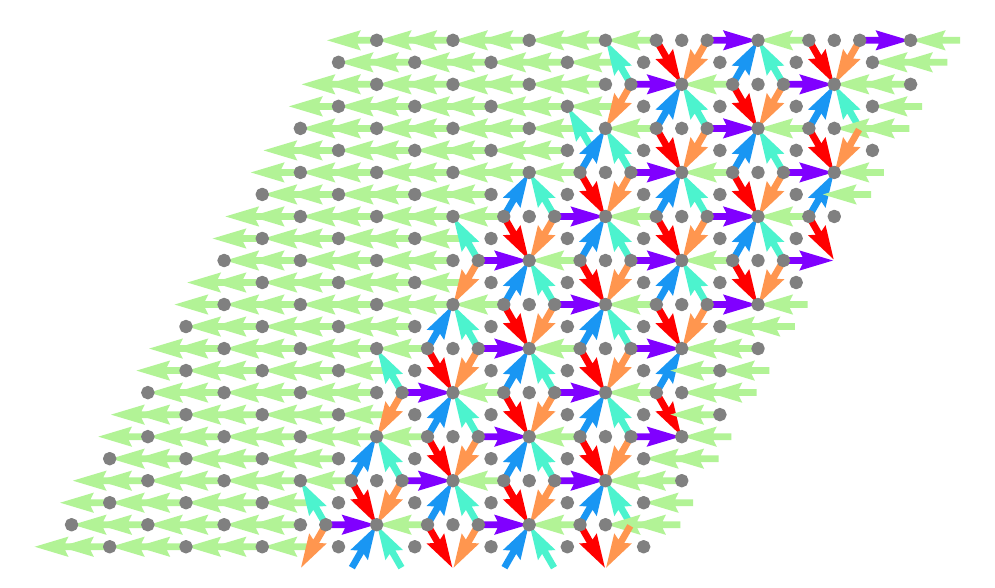}
\end{center}
\caption{
Phase separation in the SF6C model with $Q=+5$ ions at $\rho=\frac{15}{36}$. 
}
\label{fig:sf6cphasesep}
\end{figure}

%
In the SF6C model with $Q=1$, as $\rho$ is increased above $\frac{1}{2},$ we find phase separation between ions and the flower-petal phase.
In contrast, when $Q=5$, the petals are destroyed by increasing the ion concentration. 
Upon decreasing the ion concentration below $\rho=\frac{1}{2}$, the $Q=1$ system transitions to the cell wall phase, while the $Q=5$ system exhibits phase separation between the true ionic crystal and the topological crystal, Fig.~\ref{fig:sf6cphasesep}. 
%

\section{Conclusions}

In this work, we introduced and studied the phase behavior of solute-frustrated $q$-clock models, in which mobile solutes create frustration and can assemble into non-trivial phases. 
The solutes can crystallize, and this spin-mediated crystallization can be rationalized by mapping the solute-frustrated $q$-clock model onto an antiferromagnetic Ising model with couplings that are renormalized according to an effective dielectric theory. 
For low charge and a hexagonal lattice, the solutes can assemble into walls that enclose domains of spins that are analogous to microemulsions. 
The spin domains have non-trivial topology, such that the center of each spin domain hosts a vortex whose topological charge is neutralized by the surrounding solute boundary. 
We also developed a pair approximation-level mean field theory that demonstrates that a ferromagnetic phase is impossible in systems with high ionic charge and concentrations above a critical value, $\rho_c$, and predicts the tendency for neighboring spins to anti-align with one another.

Our results also suggest several directions for future investigation.
The phase diagrams reported here implicitly map out where solvent fluctuations are largest: the frustrated domain phases, with their divergent low-temperature susceptibilities, are natural candidates for the frustration-enhanced polarization fluctuations known to accelerate electron transfer~\cite{remsing2015,bhullar2021}, while the isolated paramagnetic phase suggests that this enhancement should collapse above the isolation thresholds $\rho\geq\frac{1}{2}$ (square) and $\rho\geq\frac{2}{3}$ (hexagonal). 
Since $Q$ maps onto a combination of the cation charge and the Debye screening length, which are experimentally tunable~\cite{stillinger1983,wu1992,woo1995,woo1996,limmer2015,bozym2015,ludwig2018,thenuwara2016}, these predictions are testable. 
The SF$q$C models also suggest that the type of annealed frustration studied here could be leveraged as a general strategy for topological defect-driven self assembly, analogous to vortex lattices in superconductors~\cite{abrikosov1957}, skyrmion crystals~\cite{muhlbauer2009}, and colloidal assembly in nematic liquid crystals~\cite{poulin1997,musevic2006}. 
Dynamics in SF$q$C models would also be of interest, including relaxation of the structure factor that should be governed by $d$ and $\xi$~\cite{levin1993,mundy1992} and ion transport involving collective dynamics~\cite{roy2017,remsing2017}.
Future work should also examine the impact of long-range ion-ion interactions, which we anticipate will convert macroscopic phase separation into microphase separation~\cite{grousson2000,grousson2001,deem1994a}.

\appendix

\section{Approximate Theories for Crystallization Transitions in SF$q$C models}
\label{sec:mapping}

\subsection{Dielectric Theory}

In the crystalline phase, spin-spin interactions are negligible, $\Hb^{\rm SS}\approx0$, because spin-spin contacts are rare and their contribution small compared to ion-ion and spin-ion contacts. 
Therefore, we approximate the total Hamiltonian as
\begin{equation}
\Hb \approx \Hb^{\rm SI} + \Hb^{\rm II} = -KQ \sum_{\avg{i,j}} (1-t_i)t_j \cos(\theta_i-\phi_{ij}) + LQ^2 N_{\rm II},
\end{equation}
where $N_{\rm II}=\sum_{\avg{i,j}}t_i t_j$ is the number of ion-ion contacts.
The spin-ion interaction can be rewritten in the form of a Zeeman-type interaction by noting that the sum over the nearest neighbors of $i$ is a local field on site $i$ from its neighbors, 
\begin{equation}
-\sum_i (1-t_i) \hbl_i\cdot\mathbf{S}_i,
\end{equation}
where 
\begin{equation}
\hbl_i = KQ \sum_{\mathbf{\gamma}} (1-t_{i+\mathbf{\gamma}}) \hat{\mathbf{\gamma}} \equiv KQ \mathbf{m}_i
\end{equation}
and $\hat{\mathbf{\gamma}}$ is the unit vector from $j$ to $i$. 
We converted the sum over $j$ to a sum over all the nearest neighbor unit vectors, $\mathbf{\gamma}$, such that a neighboring site is defined by the index $i+\mathbf{\gamma}$.
The local field on spin $i$, $\hbl_i$, is nonzero when at least one neighbor is a spin and at least one neighbor is an ion and the spins do not cancel upon summing. 
In a sense, one can consider $\hbl_i$ to represent the unbalanced force on $i$ arising from its neighbors. 
We can then compute an effective Hamiltonian by integrating out the spin degrees of freedom, replacing the spin-ion Hamiltonian by a free energy or potential of mean force, $\Hb_{\rm eff} = \Hb^{\rm II} + F_{\rm S}$. 
We can compute the potential of mean force by considering each spin to be independent and factorizing, such that $F_{\rm S} = \sum_i (1-t_i) f_q(\hbl_i)$, where $f_q(\hbl_i)$ is the free energy of a single spin,
\begin{equation}
f_q(\hbl) = -\kT \ln \Zb_q(\hbl).
\end{equation}
The partition function of a single spin is given by an average over the spin orientations,
\begin{equation}
\Zb_q(\hbl) = \frac{1}{q} \sum_{n=0}^{q-1} e^{\beta \hbl \cdot \mathbf{S}_n} \equiv \avg{e^{\beta \hbl \cdot \mathbf{S}_n}}_0,
\end{equation}
which defines $\avg{\cdots}_0$ as the average over spin orientations for a single spin. 
We can then approximate the free energy by expanding to second order in the reciprocal temperature,
\begin{align}
f_q(\hbl)&=-\kT\ln\avg{e^{\beta \hbl \cdot \mathbf{S}_n}}_0 \nonumber \\
&\approx -\hbl\cdot \avg{\mathbf{S}_n}_0 - \frac{\beta}{2} \avg{(\hbl\cdot \mathbf{S}_n)^2}_0.
\end{align}
The averages can be computed, $\avg{\mathbf{S}_n}_0=0$ by symmetry and $\avg{S_\alpha S_\sigma}=\frac{1}{2}\delta_{\alpha\sigma}$, where $\delta_{\alpha\sigma}$ is the Kronecker delta function and $\alpha$ and $\sigma$ can be $x$ or $y$. 
The free energy for a single spin is then given by
\begin{equation}
f_q(\hbl) = -\frac{\beta}{4}\len{\hbl}^2 = -\frac{\beta K^2 Q^2}{4} \len{\mathbf{m}}^2,
\end{equation}
and the potential of mean force is
\begin{equation}
F_{\rm S} = -\frac{\beta K^2 Q^2}{4}  \sum_i (1-t_i) \len{\mathbf{m}_i}^2.
\end{equation}
Combining this will the ion-ion interaction gives the effective Hamiltonian,
\begin{equation}
\Hb_{\rm eff} = LQ^2 N_{\rm II} - \frac{\beta K^2 Q^2}{4}  \sum_i (1-t_i) \len{\mathbf{m}_i}^2.
\end{equation}
The first term in the Hamiltonian is the direct or bare ion-ion interaction. 
The second term, the temperature-dependent potential of mean force, reduces the energy and can be likened to an indirect or solvation free energy-like term. 
We can further simplify the effective Hamiltonian by noting that $\sum_i (1-t_i) \len{\mathbf{m}}^2$ counts the number of unbalanced spins --- spins that have at least one spin and one ion neighbor and the spins do not cancel upon summing over all neighbors. 
We can try to estimate this number by noting that the crystalline phase has no ion-ion and no spin-spin contacts, and creating ion-ion contacts creates spin-spin contacts. 
As a result, the number of unbalanced spins is proportional to the number of ion-ion contacts, 
\begin{equation}
\sum_i (1-t_i) \len{\mathbf{m}_i}^2\approx a N_{\rm II},
\end{equation}
where $a$ is the number of unbalanced spins per ion-ion contact, which depends on the geometry of the lattice and the arrangement of the ions. 
Using this approximation, we can write the effective Hamiltonian in a lattice gas form,
\begin{equation}
\Hb_{\rm eff} = \para{LQ^2 - \frac{\beta K^2 Q^2 a}{4}} N_{\rm II} \equiv U_{\rm eff}(T) N_{\rm II}. 
\end{equation}
Because the effect of the spins is to reduce the magnitude of the ion-ion interactions, i.e. screening, we can write the effective interaction as a Coulomb-like screening by defining an effective dielectric constant,
$U_{\rm eff}(T) = U_0/\varepsilon(T)$ with $U_0 = LQ^2$ the bare ion-ion interaction. 
The resulting dielectric constant is
\begin{equation}
\varepsilon(T) = \para{1-\frac{K^2 a}{4L\kT}}^{-1} \sim 1+\frac{K^2a}{4L \kT},
\end{equation}
which has the Curie-Debye form expected for independent dipoles, $\varepsilon-1\sim1/T$. 
We can now map the effective Hamiltonian to the antiferromagnetic (AFM) Ising model
if we let $N=\sum_i t_i$ and $\mu$ be the chemical potential, 
\begin{equation}
\mathcal{H}_{\rm eff}-\mu N=E_0-H\sum_{i}S_i-J\sum_{\langle i,j\rangle}S_iS_j,
\end{equation}
such that $E_0=-\frac{\ell^2\mu}{2}+\frac{U_{\rm eff}\ell^2z}{8}$ is an ignorable zero-point energy, $H=\frac{\mu}{2}-\frac{U_{\rm eff}z}{4}$ is an external magnetic field, $J=-\frac{U_{\rm eff}}{4}$ is the spin-spin coupling constant, and $z$ is the coordination number~\cite{leeyang1952}. 
Because $U_{\rm eff}>0$, $J<0$, and this model represents an antiferromagnetic (AFM) Ising model. 
For the bare lattice gas model, the critical temperature is given by $\kT_c^{(0)} = cU = cLQ^2$, where $c$ depends on the lattice. 
Using our mapping, we can estimate the critical temperature for the crystalline phase using the effective Hamiltonian,
\begin{equation}
\kT_c^{\rm eff} = \kT_c^{(0)} \brac{1-\frac{K^2 a}{4L\kT_c^{\rm eff}}}.
\end{equation}
This can be rearranged to produce a quadratic equation for the critical temperature,
\begin{equation}
\para{\kT_c^{\rm eff}}^2 - \kT_c^{\rm (0)}\kT_c^{\rm eff} - \frac{\kT_c^{\rm (0)} K^2 a}{4L}= 0,
\end{equation}
whose solution is
\begin{equation}
\kT_c^{\rm eff} = \frac{\kT_c^{\rm (0)}}{2}\brac{1+\para{1-\frac{a K^2}{\kT_c^{\rm (0)}L}}^{1/2}}.
\label{eq:quadsol}
\end{equation}
Expanding the square root to first order shows that, to leading order, screening in the SF$q$C model lowers the critical temperature from that of the bare ion-ion system,
\begin{equation}
\kT_c^{\rm eff} \approx \kT_c^{\rm (0)} - \frac{aK^2}{4L}. 
\end{equation}
We can show the reduction in the critical temperature by plotting the order parameter as a function of temperature for the lattice gas and for the SF$q$C models, Fig.~\ref{fig:crystop}. 
The location of the transition is indeed reduced in the SF$q$C model. 
If we replace the reduced temperature $\kT/J$ for the lattice gas by the reduced temperature of the SF$q$C model, $\kT/J_{\rm eff}$, we find that the order parameter curves overlap. 
The relative scaling is given by the dielectric constant, $J/J_{\rm eff}=\varepsilon(T)$. 
To estimate $\varepsilon(T)$, we need to estimate $a$, which can be done for extreme cases. 
If we consider a domain wall within the crystalline phase, this creates two unbalanced spins per ion-ion contact on a square lattice, $a=2$. 
If we consider random configurations, 
\begin{equation}
a=\frac{z\rho(1-\rho)^2}{z\rho^2/2 - N^{\rm Xstal}_{\rm II}/N},
\end{equation}
which is $a=1$ for the SF4C model; $N^{\rm Xstal}_{\rm II}$ is the number of ion-ion contacts in the crystal, which is zero for the SF4C crystal and the Crystalline I phase of the SF6C model, but nonzero for the Crystalline II phase of the SF6C model. 
Finally, we note that effective charges can be obtained from
\begin{equation}
Q_{\rm eff}^2(T) = \frac{Q^2}{\varepsilon(T)},
\end{equation}
such that 
\begin{equation}
Q_{\rm eff}(T) \sim Q-\frac{aK^2Q}{8L\kT},
\end{equation}
which at the critical temperature is
\begin{equation}
Q_{\rm eff}(T_c) \sim Q-\frac{aK^2}{8cLQ}.
\end{equation}
The charge also determines whether or not a transition can occur, if we consider the argument of the square root in Eq.~\ref{eq:quadsol}.
For a square lattice, the argument becomes negative for $\len{Q}$ smaller than values between 1.3-1.8 (using Onsager's $c\approx 0.5673$), consistent with the absence of a crystallization transition in SF4C for $Q=1$. 
The range of charge values comes from the spread in $a$ given by domain wall and random mixing estimates.
For a hexagonal lattice, the same estimates suggest that there is no transition when $\len{Q}$ is less than 2.8-3.45, again consistent with the absence of a crystallization transition in the corresponding $Q=1$ system. 

\subsection{Isolated System Approximation}

\begin{figure}
\begin{center}
\includegraphics[width=0.45\textwidth]{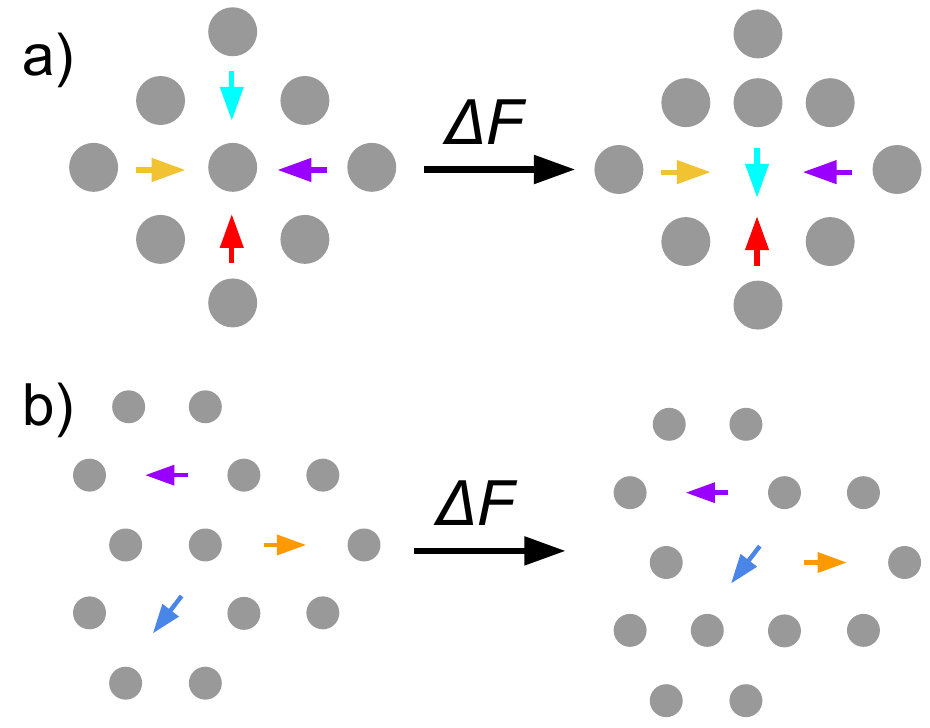}
\end{center}
\caption{
Defect creation process in (a) the SF4C model at $\rho=\frac{1}{2}$ and (b) the SF6C model at $\rho=\frac{2}{3}$.
}
\label{fig:latticedefectdiagrams}
\end{figure}

%
A more numerically exact argument may be found by considering the thermodynamic penalty of melting the lattice with and without ions in an isolated system.
Consider a lattice gas on a square lattice, and consider an isolated system in which the ions form a perfect lattice.
We consider the free energetic penalty for creating a defect in this lattice whereby we move the central ion up one unit, shown in Fig.~\ref{fig:latticedefectdiagrams}a.
In this case, $\Delta F^{\rm LG}=3LQ^2,$ as the central ion starts with zero nearest-neighbor interactions and ends with 3. Now, let us consider the SF4C model, adding back in the spins and again consider the penalty for creating a defect.
We will define this to be $\Delta F^{\rm SF4C}=3LQ_{\rm eff}^2,$ where $Q_{\rm eff}$ is now the effective charge which would provide the same energetic penalty of creating the defect.

To calculate $Q_{\rm eff},$ we can explicitly compute $\Delta F^{\rm SF4C}$.
First, we observe that the partition function of the perfect lattice is $Z_0=\sum_{\nu}1=4^4=256,$ since we have 4 spins which each can take 4 different orientations. 
For the case of the defect, we can calculate $E(\nu)$ by evaluating $\mathcal{H}$ for each state.
Each state has an ion-ion energy of $3LQ^2,$ while the different configurations will have different spin-ion and spin-spin energies.
Specifically, there are four spin-ion pairs and three spin-spin pairs whose energies must be considered.
We then do this for each of the 256 microstates, which is a small enough number that the results may be computed analytically.
From here, we simply compute $Z_f(T)=\sum_{\nu}e^{-\beta E(\nu)},$ though for convenience, it is easier to condense all terms with the same energy such that $Z_f(T)=\sum_E g(E)e^{-\beta E},$ where $g(E)$ is the number of states with energy $E$.
Naturally, then, 
\begin{equation}
\Delta F^{\rm SF4C}=-\kb T_c^{(0)}\ln\left(\frac{Z_f\left(T_c^{(0)}\right)}{256}\right),
\end{equation}
which yields 
\begin{equation}
Q_{\rm eff}=\sqrt{-\frac{\kb T_c^{(0)}}{3}\ln\left(\frac{Z_f\left(T_c^{(0)}\right)}{256}\right)}.
\end{equation}
Notice that as before, the effective charge depends on the temperature, and we will again use $T_c^{(0)}$ as our input temperature.
Doing this calculation yields $Q_{\rm eff}\simeq 4.940,$ very close to the simulated $4.927$.
Similarly, we can calculate the temperature shift as 
\begin{equation}
\frac{T_c^{(0)}}{T_c^{\rm eff}}=\left(\frac{Q}{Q_{\rm eff}}\right)^2,
\end{equation}
which yields 1.024, very close to the simulated 1.03.

\begin{figure}
\begin{center}
\includegraphics[width=0.45\textwidth]{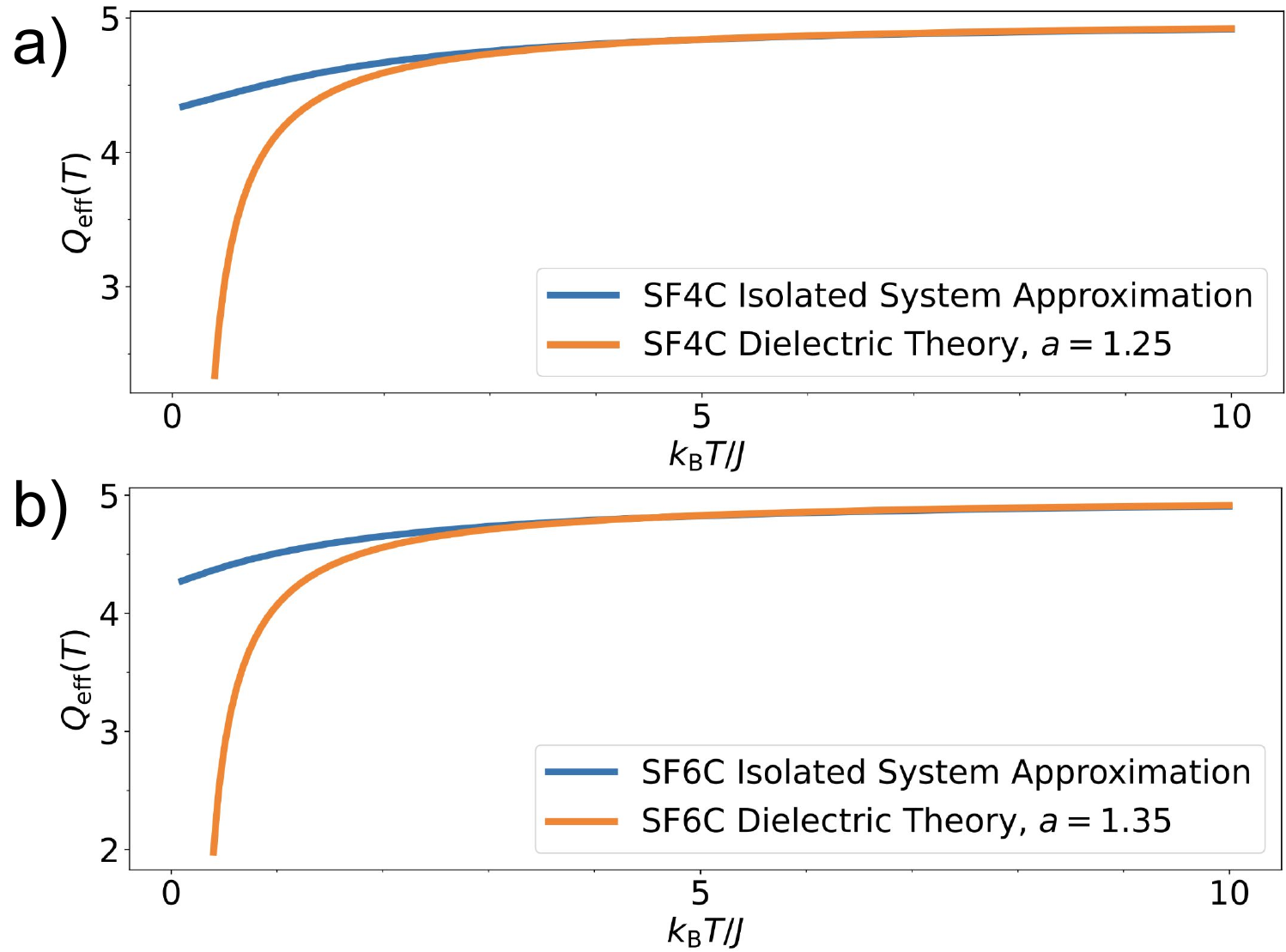}
\end{center}
\caption{
Effective charge in the (a) SF4C model at $\rho=\frac{1}{2}$ and (b) SF6C model at $\rho=\frac{2}{3}$ as a function of temperature using both approximation methods presented.
}
\label{fig:effectivecharge}
\end{figure}

We can do a similar analysis for the SF6C model.
We only consider the case of $\rho=\frac{2}{3}$ since the spins are not isolated for $\rho=\frac{1}{3}$, which makes our method intractable.
The specific configuration being considered is shown in Fig.~\ref{fig:latticedefectdiagrams}b.
The free energy will be 
\begin{equation}
\Delta F^{\rm SF6C}=2LQ_{\rm eff}^2=-\kb T_c^{(0)}\ln\left(\frac{Z_f\left(T_c^{(0)}\right)}{216}\right),
\end{equation} where our partition function is summed over the 6 states of each of the 3 spins considered.
This yields a $Q_{\rm eff}\simeq 4.898,$ similar to the simulated 4.888, and with a temperature shift of 1.042, similar to the simulated 1.046.

Both approximation methods lead to very similar conclusions.
For an appropriate choice of $a$, the number of unbalanced spins per ion-ion contact, we find that the methods numerically agree at near the critical temperature and at higher temperatures, as shown in Fig.~\ref{fig:effectivecharge}. 
We also see that both theories correctly describe that the ionic charges are screened by the spins, leading to a slightly decreased energetic penalty for melting the lattice.

\section{Mathematical Development of a mean field Theory}
\label{sec:mft}

We now analyze our model using mean field theory. 
In the limit without ions, Goswami \etal~found that mean field theory successfully describes the phase behavior of $q$-clock models, $q\geq 5$. 
Specifically, they find two phase transitions, a BKT-like transition to a phase with $U(1)$ symmetry and a spontaneous symmetry breaking transition to a phase with $\mathbb{Z}_q$ symmetry~\cite{goswami2024}. 
We shall adapt this mean field theory by adding in ions. 
Our mean field is as follows---a sea of charged spins with spin $(1-\rho)\sigma$, with $\sigma=\sigma_x+i\sigma_y$ being the desired quantity, and charge $\rho Q$, $Q$ being the ionic charge. 
Traditional MFT, considering only a single site interacting with a mean field, only gives a trivial result, as the interaction from the uniform charge cancels out, leaving only a weakened spin-spin interaction. 
As a result, we get exactly the same phases as in the unfrustrated $q$-clock model, though with different transition temperatures. 
Thus, we apply a pair approximation~\cite{bethe1935,oguchi1955,weeks1975pair,goswami2024}. 
Consider two lattice sites, 0 and 1, with values $t_0$, $S_0=e^{i\theta_0}$, $t_1$, and $S_1=e^{i\theta_1}$, surrounded by a mean field; recall that $t_i=0$ if site $i$ is a spin and 1 if it is an ion. 
We model the interaction between these two sites exactly while using the mean field for all other interactions. 
We call the spins of the $z-1$ mean field neighbors of site zero $S_{i,0}$ and likewise those of site one $S_{i,1}$---note that $z$ is the coordination number of the lattice. 
To begin, we will need to write a two-site Hamiltonian. 
First, we have the pure spin-spin Hamiltonian---note that interactions with the mean field must be halved to avoid double counting. 
Then, we write the spin-ion contribution, which includes a spin interacting with the charged part of the mean field as well as a spin interacting exactly with an ionic neighbor in the pair. 
Moreover, it also includes an ion interacting with the spins of the mean field. 
Finally, we write the ion-ion contribution. 
We get the following for our two-site Hamiltonian. 
\begin{widetext}
\begin{equation}
\mathcal{H}^{\mathrm{SS}}_{\mathrm{MF}}=-J\real\left((1-t_0)(1-t_1)S_0S_1^*+\frac{1}{2}\left[\sum_{i=1}^{z-1}(1-t_0)S_0S_{i,0}^*+\sum_{i=1}^{z-1}(1-t_1)S_1S_{i,1}^*\right]\right)\\
\end{equation}
\begin{equation}
\mathcal{H}^{\mathrm{SI}}_{\mathrm{MF}}=-KQ\left(-(1-t_0)t_1S_0^x+(1-t_1)t_0S_1^x+\frac{1}{2}\left[\rho (1-t_0)S_0^x-\rho (1-t_1)S_1^x-t_0\sigma_x+t_1\sigma_x\right]\right)\\
\end{equation}
\begin{equation}
\mathcal{H}^{\mathrm{II}}_{\mathrm{MF}}=LQ^2\left(t_0t_1+\frac{1}{2}\left[(z-1)\rho t_0+(z-1)\rho t_1\right]\right)\\
\end{equation}
\begin{equation}
\mathcal{H}_{\mathrm{MF}}=\mathcal{H}^{\mathrm{SS}}_{\mathrm{MF}}+\mathcal{H}^{\mathrm{SI}}_{\mathrm{MF}}+\mathcal{H}^{\mathrm{II}}_{\mathrm{MF}}
\end{equation}
\end{widetext}
Now, our system has four possible states: both sites are spins, only site 0 is a spin, only site 1 is a spin, and both sites are ions. 
Using the above Hamiltonian, we can write a different Hamiltonian for each of these states. 
Moreover, these Hamiltonians will depend on $\sigma_x$ and $\sigma_y$, our mean field spin, as well as $k_0$ and $k_1$, which are the individual spin orientations of the two sites. 
We follow a similar method as described by Goswami \etal~\cite{goswami2024}. 
Finally, this allows us to write the partition function $Z(\sigma_x,\sigma_y)$, from which $F(\sigma_x,\sigma_y)=-\kb T\ln(Z(\sigma_x,\sigma_y))$ is our Helmholtz free energy. 
\begin{widetext}
\begin{equation}
\begin{split}
h_{0,0}(\sigma_x,\sigma_y,k_0,k_1)=&-J\left\{-\left(z-1\right)\left(1-\rho\right)\left(\sigma_x^{2}+\sigma_y^{2}\right)+\cos\left(\frac{2\pi}{q}\left(k_0-k_1\right)\right)\right\}\\&-J\left\{\left(z-1\right)\left(1-\rho\right)\left(\sigma_x\left[\cos\left(\frac{2\pi k_0}{q}\right)+\cos\left(\frac{2\pi k_1}{q}\right)\right]+\sigma_y\left[\sin\left(\frac{2\pi k_0}{q}\right)+\sin\left(\frac{2\pi k_1}{q}\right)\right]\right)\right\}\\&-\frac{KQ\rho}{2}\left(\cos\left(\frac{2\pi k_0}{q}\right)-\cos\left(\frac{2\pi k_1}{q}\right)\right),
\end{split}
\end{equation}
\begin{equation}
\begin{split}
h_{0,1}\left(\sigma_x,\sigma_y,k_0\right)=&-J\left\{-\frac{\left(z-1\right)}{2}\left(1-\rho\right)\left(\sigma_x^{2}+\sigma_y^{2}\right)+\left(z-1\right)\left(1-\rho\right)\left(\sigma_x\cos\left(\frac{2\pi k_0}{q}\right)+\sigma_y\sin\left(\frac{2\pi k_0}{q}\right)\right)\right\}\\&+KQ\left(\cos\left(\frac{2\pi k_0}{q}\right)-\frac{\sigma_x\left(1-\rho\right)}{2}\right)+LQ^{2}\left(\frac{\left(z-1\right)\rho}{2}\right),
\end{split}
\end{equation}
\begin{equation}
\begin{split}
h_{1,0}\left(\sigma_x,\sigma_y,k_1\right)=&-J\left\{-\frac{\left(z-1\right)}{2}\left(1-\rho\right)\left(\sigma_x^{2}+\sigma_y^{2}\right)+\left(z-1\right)\left(1-\rho\right)\left(\sigma_x\cos\left(\frac{2\pi k_1}{q}\right)+\sigma_y\sin\left(\frac{2\pi k_1}{q}\right)\right)\right\}\\&-KQ\left(\cos\left(\frac{2\pi k_1}{q}\right)-\frac{\sigma_x\left(1-\rho\right)}{2}\right)+LQ^{2}\left(\frac{\left(z-1\right)\rho}{2}\right),
\end{split}
\end{equation}
\begin{equation}
h_{1,1}\left(\sigma_x,\sigma_y\right)=LQ^{2}\left(1+\left(z-1\right)\rho\right).
\end{equation}
\begin{equation}
\begin{split}
Z(\sigma_x,\sigma_y)&=\sum e^{-\beta\mathcal{H}_{\mathrm{MF}}}=\sum_{i,j\in \{0,1\}}\sum_{k_0,k_1=0}^{q}P(t_0=i,t_1=j)e^{-\beta h_{i,j}(\sigma_x,\sigma_y,k_0,k_1)}\\&=\sum_{k_0,k_1=1}^{q}(1-\rho)^2e^{-\beta h_{0,0}(\sigma_x,\sigma_y,k_0,k_1)}+\sum_{k_0=1}^{q}\rho(1-\rho)e^{-\beta h_{0,1}(\sigma_x,\sigma_y,k_0)}\\&+\sum_{k_1=1}^{q}\rho(1-\rho)e^{-\beta h_{1,0}(\sigma_x,\sigma_y,k_1)}+\rho^2e^{-\beta h_{1,1}(\sigma_x,\sigma_y)}.
\end{split}
\end{equation}
\end{widetext}
From the free energy, we are able to get two quantities of interest. 
First, to get the net magnetization, we find the value of $\sigma$ at the minima of $F$, though we are usually only interested in $|\sigma|$. 
We can also compute the nearest-neighbor spin-spin (NNSS) correlation function, since we have treated the interaction between two spins exactly. 
We will modify the method used by Goswami \etal~\cite{goswami2024} for this calculation. 
The NNSS is defined as $C(1)=\langle \cos(\theta_i-\theta_j)\rangle$ where we average over nearest neighbor spin pairs $i$ and $j$. 
However, we also need to account for the fact that certain NN pairs consist of a spin and an ion or even an ion and an ion. 
These should not be included in the averaging process, since we only want contributions from spin-spin pairs. 
Thus, we only consider configurations with two spins. 
Therefore, we find 
\begin{widetext}
\begin{equation}
C(1)=\langle\cos(\theta_i-\theta_j)\rangle = \frac{\sum_{k_0,k_1=1}^{q}\cos\left(\frac{2\pi}{q}(k_0-k_1)\right)(1-\rho)^2e^{-\beta h_{0,0}(\sigma_x,\sigma_y,k_0,k_1)}}{\sum_{k_0,k_1=1}^{q}(1-\rho)^2e^{-\beta h_{0,0}(\sigma_x,\sigma_y,k_0,k_1)}}.
\end{equation}
\end{widetext}
\noindent Note that in order to compute this, we also need to know $\sigma_x$ and $\sigma_y$, and thus the magnetization. 
There are a few limitation to our mean field approach. 
The mean field smears out ionic charges and therefore cannot account for crystallization of ion sites, which is necessary for phases like topological crystals. 
Rotational symmetry is broken in the mean field theory because of the explicit pair interaction, and we also assume that the spin occupation probability is purely statistical. 
Nevertheless, the MFT does produce qualitatively accurate and physically useful predictions for the state points of interest. 

\begin{acknowledgements}
We acknowledge the Office of Advanced Research Computing (OARC) at Rutgers, The State University of New Jersey for providing access to the Amarel cluster and associated research computing resources that have contributed to the results reported here.
\end{acknowledgements}

\bibliography{SFqC}

\end{document}